\documentclass[fleqn,10pt]{SelfArx}
\usepackage{graphicx}
\usepackage{amsmath, mathrsfs, amsthm}
\usepackage{amsfonts}
\usepackage{amssymb}
\usepackage{wasysym}
\usepackage{upgreek}
\usepackage{subfigure}
\usepackage{color}
\usepackage{multirow}
\usepackage{multicol}
\usepackage{tabularx}
\usepackage{float}
\usepackage{makecell}
\usepackage[flushleft]{threeparttable}
\usepackage{todonotes}
\usepackage{dutchcal}
\usepackage{moresize}

\newcommand\abss[1]{\lvert#1\rvert}

\newcommand\curve{\boldsymbol{\varphi}}
\newcommand\curvep{\boldsymbol{\varphi}^{\prime}}
\newcommand\curvepp{\boldsymbol{\varphi}^{\prime\prime}}
\newcommand\curvedp{\boldsymbol{\dot{\varphi}}^{\prime}}
\newcommand\curved{\boldsymbol{\dot{\varphi}}}
\newcommand\curvedd{\boldsymbol{\ddot{\varphi}}}

\definecolor{myblue}{RGB}{0, 0, 176}

\theoremstyle{definition}

\theoremstyle{definition}

\theoremstyle{theorem}

\theoremstyle{corollary}

\theoremstyle{proposition}

\theoremstyle{lemma}

\usepackage{lipsum} 

\definecolor{color1}{RGB}{0,0,90} 
\definecolor{color2}{RGB}{0,20,20} 

\usepackage{hyperref} 
\hypersetup{hidelinks,colorlinks,breaklinks=true,urlcolor=color2,citecolor=color1,linkcolor=color1,bookmarksopen=false,pdftitle={Title},pdfauthor={Author}}

\JournalInfo{Arxiv File} 
\Archive{Article submitted to Marine Structures - December, 2024} 

\PaperTitle{On the use of an advanced Kirchhoff rod model to study mooring lines} 

\Authors{Bruno A. Roccia\textsuperscript{1}*, Hoa T. Nguyen\textsuperscript{1}, Petter Veseth\textsuperscript{2}, Finn G. Nielsen\textsuperscript{1}, Cristian G. Gebhardt\textsuperscript{1}} 
\affiliation{\textsuperscript{1}\textit{Bergen Offshore Wind Centre and Geophysical Institute, University of Bergen, Norway}} 
\affiliation{\textsuperscript{2}\textit{Equinor ASA, Bergen, Vestland, Norway}} 
\affiliation{*\textbf{Corresponding author}: bruno.roccia@uib.no} 

\newcommand{\keywordname}{Keywords} 

\Abstract{In this work, we investigate the application of an advanced nonlinear torsion- and shear-free Kirchhoff rod model, enhanced with a penalty-based barrier function (to simulate the seabed contact), intended for studying the static and dynamic behavior of mooring lines. The formulation incorporates conservative and non-conservative external loads, including those coming from the surrounding flow (added mass, tangential drag, and normal drag). To illustrate the favorable features of this model, we consider some key scenarios such as static configurations, pulsating force applications at the fairlead, and fluid-structure interaction between mooring lines and the surrounding flow. Verification against well-established solutions, including catenary configurations and OpenFAST simulations, shows excellent accuracy in predicting mooring line responses for a floating offshore wind turbine. 
Among the most important results, we can mention that under normal pulsating loads at the fairlead, the mooring line exhibits a transition from a drag-dominated regime at low frequencies to an added-mass-dominated regime at higher frequencies. Furthermore, tangential forcing at the fairlead reveals a strong coupling between axial and bending dynamics, contrasting with normal forcing scenarios where axial dynamics remain largely unaffected. These findings underscore the potential of the proposed approach for advanced mooring line simulations.}

\begin{document}

\flushbottom 

\maketitle 

\tableofcontents 

\thispagestyle{empty} 


\section{Introduction}\label{IntroSec}

Currently, there is a global awareness about the importance of achieving a society that meets its needs in a more sustainable way. Any serious attempt to pursue this goal undoubtedly requires to focus on environment friendly energy resources. Along this path, floating offshore wind turbines (FOWTs) have recently become one of the most attractive technologies to partially meet the global energy demand \cite{GWEC2023}.


The development, manufacturing, and installation of high-power floating offshore wind farms introduce new challenges for the offshore wind industry, \textit{particularly in the design and analysis of mooring systems} \cite{yang2022}. The dynamic behavior of these structural components is influenced by numerous environmental loads and physical factors, including wind, wave, and current forces; buoyancy and cable weight; interactions with the seafloor; snap loads; and the mechanical impedance of the line, among others \cite{davidson2017}.

When modeling mooring systems for offshore applications in general, including FOWTs, several feasible approaches can be employed, namely static, quasi-static (QS), frequency-domain (FD), and time-domain analyses. The former are typically conducted to perform predesign studies of mooring lines, where the line pretension is specified to achieve the desired draft of the structure \cite{MA201985}. At this stage, the most common approaches rely on two- and three-dimensional elastic catenary equations \cite{MA201985,impollonia2011,Gunnar2024} and simplified lumped-parameter models (LPMs) \cite{roccia2020cable}. Quasi-static solvers are widely used for FOWTs due to their computational efficiency and availability in open-source software, such as MAP++ (Mooring Analysis Program) integrated within OpenFAST \cite{masciola2013}. The primary limitation of QS methods is the neglect of hydrodynamic and inertial forces, leading to a general underestimation of loads \cite{matha2011non}. Although quasi-static approaches were commonly employed during the early development of mooring analysis and remain useful \cite{hall2024generalized} for rapid estimates, the majority of modern mooring designs now rely on dynamic analysis \cite{MA201985}.

In the realm of dynamic analyses, frequency-domain methods are widely employed due to their computational efficiency, particularly when quick solutions are required. FD analysis is also commonly used for calculating fatigue damage, as it allows for the evaluation of a large number of load cases. In FD analysis, the system response is typically expressed as a combination of static and frequency-dependent components, thus enabling the application of the superposition principle across different frequencies. However, FD-based solvers are limited to linear analysis \cite{he2021time} and decouple the mooring line from the floating structure, solving the motion of the structure independently of the tension in the mooring line \cite{MA201985}. 

Regarding time-domain analyses, we can distinguish between lumped-parameter models and distributed-parameter models (DPMs). Numerous two- and three-dimensional implementations of LPMs are available in the literature, and these models are widely adopted within the industry. LPMs can capture critical phenomena such as seabed reaction \cite{wilhelmy1981,nakajima1982}, clump weights \cite{khan1986}, seabed friction \cite{devries2018}, and hydrodynamic loads \cite{hall2015,hermawan2020}. Hall and Goupee \cite{hall2015} developed an open-source LPM for both standalone and coupled simulations, which has been officially integrated into OpenFAST for aero-hydro-elastic simulations of floating offshore wind turbines (FOWTs). Other variants of LPMs have also been implemented and successfully applied in marine engineering mooring analysis \cite{touzon2020,hermawan2020}. For DPMs, a wide range of contributions can be found in the literature, ranging from models that neglect bending stiffness \cite{aamo2000}, to those incorporating bending and torsion \cite{buckham2004}, high-order finite element (FE) formulations \cite{escalante2011}, vector form intrinsic finite element (VFIFE) approaches \cite{zhang2022}, mixed FE formulations \cite{montano2007}, and local discontinuous Galerkin FE methods \cite{palm2017}. Zhong et al. \cite{zhong2024c} recently compared several of these mooring line models in the context of hydrodynamic performance of FOWTs. Additionally, more advanced approaches have been employed to model mooring lines subjected to large rotations and displacements, including geometrically exact beam formulations \cite{quan2020}, quaternion-based models \cite{zupan2013}, models combining NURBS curves with quaternions and an isogeometric collocation method \cite{weeger2017}, and Cosserat rod models \cite{tschisgale2020,martin2021}.

Regarding software packages, a wide range of analysis tools are available for both frequency-domain and time-domain analyses, including OrcaFlex, Sima, and OpenFAST, among other commercial and in-house developments \cite{davidson2017}. OrcaFlex, for instance, supports both nonlinear time-domain analysis and linear frequency-domain analysis, utilizing a lumped-mass mooring line model \cite{Orcina2016}. In contrast, Sima, developed by SINTEF, serves as a comprehensive workbench for simulation, interpretation, and documentation of analysis results. It is specifically designed for marine operations and floating systems, using the finite element-based program RIFLEX for modeling mooring lines \cite{SIMA2023}. Finally, OpenFAST, an open-source simulation tool developed by NREL, incorporates a lumped-parameter mooring line model (MoorDyn) for dynamic simulations \cite{Hall2017Moor}. Wendt et al. \cite{wendt2016} conducted a validation and verification study of MoorDyn against wave tank measurements and the commercial software OrcaFlex, demonstrating largely equivalent results. With respect to seabed interaction, all these tools simulate seabed reaction forces using linear or nonlinear elastic springs.

Building upon the aforementioned initiatives, the contribution of this paper is twofold. First, we present a modified version of the nonlinear torsion- and shear-free Kirchhoff rod model \cite{gebhardt2021} to investigate the static and dynamic behavior of mooring lines for FOWTs. This model is enhanced by incorporating: \textit{i}) a penalty-based barrier function to simulate seabed contact, and \textit{ii}) a simplified fluid model to account for non-conservative forces from the surrounding flow, including added mass, tangential and normal drag, and buoyancy. Additionally, we provide a review of barrier functions commonly employed in computational mechanics and optimization, highlighting their applications in marine structure modeling to represent the interaction between mooring lines and the seabed. A discussion is also included on the numerical implications of the penalty parameter in relation to the touchdown point of the mooring line. Second, we present a series of numerical simulations to evaluate the behavior of our enhanced model, hereinafter referred to as \textit{ARMoor} (Advanced Rod Model for Mooring Lines), under various scenarios and loading conditions. Finally, we discuss the limitations of the proposed numerical framework.

To the best of our knowledge, the enhanced nonlinear Kirchhoff rod formulation presented in this work for static and hydro-elastic simulations of mooring lines, including seabed interaction, has not been previously addressed in the literature. As this paper incorporates several well-established theoretical and numerical methodologies, many of the results presented here will be familiar to engineers and researchers in the fields of marine structures and marine operations. Nonetheless, we believe that ARMoor holds significant potential for simulating mooring lines under complex and challenging environmental conditions. Furthermore, it offers the capability to be integrated with multibody dynamic system models to account for interactions among multiple mooring lines and the dynamics of the floater. In this regard, the contribution of this study is novel and has not yet appeared in the existing literature.

The remainder of this paper is organized as follows: In Sect. \ref{ModSec}, we detail the nonlinear Kirchhoff rod model, including its geometric foundations. Sect. \ref{resultsSec} presents numerical results to verify ARMoor against well-known solutions from the literature and established software tools in the offshore wind energy sector. In Sect. \ref{sec:limitations}, we discuss the limitations of the current approach and outline potential improvements. Finally, Sect. \ref{ConcluSec} provides concluding remarks.


\section{Model description}\label{ModSec}

The nonlinear Kirchhoff rod model adopted in this article is based on \cite{gebhardt2021} enhanced by incorporating: \textit{i}) the fluid forces coming from the surrounding flow, and \textit{ii}) a barrier function to consider a \textit{boundary} on the structure. The latter feature is extremely useful for modeling elastic members such as mooring lines in offshore engineering applications (e.g. wind turbines or oil platforms, among others) where the mooring line may interact with the seabed. We start with a brief recapitulation of the required fundamental equations that are later used for the rod formulation. Then, we introduce the barrier function by considering an augmented version of the Lagrangian $\mathcal{L}$. 

\subsection{Preliminaries}\label{subsec:pre}

The nonlinear rod considered here is transversely isotropic, disregards the energetic contribution coming from torsion, and does not exhibit shear deformation and is initially straight. In this regard, let us consider an arbitrary regular one-parameter curve $\curve(s)$ where $s\in[0,L]$ is the arc-length coordinate. Since $\curvep$ is well-defined everywhere (i.e. $\curvep\neq 0$), the unit tangent vector to $\curve$ at every point (hereafter referred as \textit{director}) is computed as $\mathbf{d}=\curvep/\abss{\curvep}$, where $(\cdot)^{\prime}$ denotes derivative with respect to the arc-length $s$ and $\abss{\cdot}$ denotes the Euclidean vector norm. It should be noted that the definition of director implies that $\mathbf{d}$ belongs to the unit sphere $S^2:=\{\mathbf{d}\in\mathbb{R}^3,\,\mathbf{d}\cdot\mathbf{d}=1\}$. Then, we introduce a configuration space that automatically disables shear deformations as follows,    
\begin{equation}\label{eq.conf.1}
    \mathcal{D}:=\{\curve\in[C^2[0,L]]^3,\,\abss{\curvep (s)}>0\}.
\end{equation}

As usual in solid mechanics, a motion of the rod is a curve of configurations parameterized by time, i.e. $\curve_t(s)=\curve(s,t)$ for $s\in[0,L]$ and $t\in[0,T]$. In this work, we consider as class of admissible motions the ones having as initial configurations,
\begin{equation}\label{eq.conf0.2}
    \curve_0(s)=s\mathbf{D},\quad\text{with}\quad \mathbf{D}=\frac{a^k\mathbf{E}_k}{\abss{a^k\mathbf{E}_k}},
\end{equation}

\noindent
where $a^k\in\mathbb{R}$ and $\mathcal{E}=\{\mathbf{E}_k\}_{k=1}^3$ stands for the canonical Cartesian basis of $\mathbb{R}^3$. Note that Latin indices $i, j, k$, and so on range from 1 to 3. In addition, Einstein’s summation is implied unless otherwise specified. 

Finally, we introduce the concept of \textit{covariant derivative}. To this end, we first define the tangent bundle associated to $S^2$ as $TS^2:\{(\mathbf{d},\mathbf{r})\in\,S^2\times\mathbb{R}^3,\mathbf{d}\cdot\mathbf{r}=0\}$. Then, the covariant derivative  of a smooth vector field $\mathbf{u}:S^2\longrightarrow TS^2$ along a vector field $\mathbf{v}:S^2\longrightarrow TS^2$ is a vector field in $TS^2$ evaluated at $\mathbf{d}$, given by,
\begin{equation}\label{eq.covariant.3}
    \nabla_{\mathbf{v}}\mathbf{u}:=\left(\mathbf{1} - \mathbf{d}\otimes\mathbf{d} \right) D\mathbf{u}\cdot\mathbf{v},
\end{equation}

\noindent
where $\mathbf{1}$ is the identity matrix, $\mathbf{1}-\mathbf{d}\otimes\mathbf{d}$ is recognized as the orthogonal projection operator in $\mathbb{R}^3$ (hereinafter denoted by $\mathcal{P}_{\mathbf{d}}^{\perp}$), and $D\mathbf{u}\cdot\mathbf{v}$ is the directional derivative of $\mathbf{u}$ along $\mathbf{v}$. In other words, the covariant derivative \eqref{eq.covariant.3} is the orthogonal projection of the standard covariant derivative in $\mathbb{R}^3$ onto the tangent space of $S^2$ \cite{oneill2006}.


\subsection{Nonlinear shear- and torsion-free rod formulation}\label{subsec:rod}

We recall from \cite{gebhardt2021} that a variational derivation of the equations of motion (EoM) can be obtained from Hamilton's principle of least action, which states that the governing equations of a mechanical system are the Euler-Lagrange equations of the following action functional, 
\begin{equation}\label{eq.hamilton.4}
\begin{split}
    \mathcal{F} &= \int_0^T \mathcal{L}(\curvep,\curved,\curvepp,\curvedp;t)\,dt \\ &=\int_0^T\left[\mathcal{T}(\curvep,\curved,\curvedp;t) - \mathcal{U}(\curvep,\curvepp;t) \right]\,dt,
    \end{split}
\end{equation}

\noindent
where $\mathcal{T}$ is the kinetic energy density of the rod, $\mathcal{U}$ is the potential energy functional of the rod, $\mathcal{L}=\mathcal{T}-\mathcal{U}$ is the Lagrangian of the system, and $\dot{(\,\cdot\,)}$ denotes derivative with respect to time. The Euler-Lagrange equations corresponding to the Kirchhoff rod are obtained by taking the variation of the functional \eqref{eq.hamilton.4} and adding the virtual work of all the external forces (gravitational forces, hydrodynamic forces, buoyancy forces, etc.), i.e.,
\begin{equation}\label{eq.weak.4}
    \overline{\delta\mathcal{F}} = \int_0^T\left[\delta\mathcal{T}(\curvep,\curved,\curvedp;t) - \delta\mathcal{U}(\curvep,\curvepp;t) -\overline{\delta\mathcal{W}}\right]\,dt = 0.
\end{equation}

Following \cite{gebhardt2021}, the strong form of the EoMs governing the space-time evolution of the Kirchhoff rod are given by, 
\begin{equation}\label{eq.EoMstrong.5}
    \mathbf{n}^{\prime} + \left( \frac{1}{\abss{\curvep}}\mathbf{d}\times\nabla_{\mathbf{d}^{\prime}}\mathbf{m}\right)^{\prime} = A_{\rho}\curvedd + \left( \frac{1}{\abss{\curvep}}\mathbf{d}\times I_{\rho}\nabla_{\dot{\mathbf{d}}}\dot{\mathbf{d}}\right)^{\prime} - \mathbf{f}_{ext},
\end{equation}

\noindent
where $A_{\rho}$ and $I_{\rho}$ are the mass per unit length and the inertia density, respectively, $\mathbf{f}_{ext}$ collects all the external generalized forces, and $\mathbf{n}=EA\boldsymbol{\epsilon}$ and $\mathbf{m}=EI\boldsymbol{\kappa}$ are the stress measures, which are conjugated with the following strain measures,
\begin{equation}\label{eq.strain.6}
    \boldsymbol{\epsilon}:=\curvep - \mathbf{d},\quad\text{and}\quad \boldsymbol{\kappa}:=\mathbf{d}\times\mathbf{d}^{\prime},
\end{equation}

\noindent
with $E$ being the elastic Young's modulus, and $A$ and $I$ being the cross-section area and moment of inertia of the rod, respectively. 

The governing equation \eqref{eq.EoMstrong.5} is complemented by a set of initial conditions (IC) and boundary conditions (BC). For the ICs, we require $\curve(0)=\curve_0$ and $\curved(0)=\mathbf{v}_0$ on $(s,t)\in[0,L]\times\{0\}$. Although we are not limited to, for simplicity, here we consider two different BCs: clamped-free and simply supported at both ends. Such BCs are detailed in Table \ref{T1}. It should be noted that $\curve_{\text{end}}=r^k\mathbf{E}_k$, with $r^k\in\mathbb{R}$, is the position vector of a spacial point in $\mathbb{R}^3$.

\begin{table*}[h!]
\centering
\caption{Boundary conditions}
\label{T1}
\begin{tabular}{ >{\arraybackslash}m{.20\linewidth} 
>{\centering\arraybackslash}m{.20\linewidth} 
>{\centering\arraybackslash}m{.40\linewidth}}
\hline\hline\vspace{0.1cm}
BC & at $s=0$ & at $s=L$ \\  
\hline\vspace{0.1cm}
\multirow{2}{*}{Clamped-free} & $\curve=\mathbf{0}$ & $\mathbf{n}+\frac{1}{\abss{\curvep}}\mathbf{d}\times\left(\nabla_{\mathbf{d}^{\prime}}\mathbf{m} - I_{\rho} \nabla_{\dot{\mathbf{d}}}\dot{\mathbf{d}}\right)=\mathbf{0}$  \\[0.1cm]
                          & $\curvep=\mathbf{D}$  & $\frac{1}{\abss{\curvep}}\mathbf{d}\times\mathbf{m}=\mathbf{0}$  \\[0.1cm] \cline{2-3}
\multirow{2}{*}{Simply-supported} & $\curve=\mathbf{0}$ & $\curve = \curve_{\text{end}}$  \\[0.1cm]
                              & $\frac{1}{\abss{\curvep}}\mathbf{d}\times\mathbf{m}=\mathbf{0}$ & $\frac{1}{\abss{\curvep}}\mathbf{d}\times\mathbf{m}=\mathbf{0}$  \\ [0.1cm]
\hline\hline
\end{tabular}
\end{table*}

Finally, the weak form associated to the strong form \eqref{eq.EoMstrong.5} can be expressed as,
\begin{equation}\label{eq.EoMweak}
    \int_0^L \delta\curve\cdot\left[ \mathcal{M}(\curvep)\hat{\nabla}_{\curved}\curved +\mathcal{B}^T(\curvep,\curvepp)\boldsymbol{\sigma} - \mathbf{f}_{ext} \right]\,ds = 0,
\end{equation}

\noindent
where $\mathbf{\mathcal{B}}$ is the linearized strain operator, $\mathcal{M}$ is the mass operator, $\boldsymbol{\sigma}:=(\mathbf{n}, \mathbf{m})^T$, and $\hat{\nabla}_{(\cdot)}(\cdot)$ is the field extension of the covariant derivative \eqref{eq.covariant.3}. The reader is referred to \cite{gebhardt2021,Nguyen2024} for a full description of such operators.


\subsection{Barrier function}\label{subsec:barrier}

\begin{figure}[h]
\centering
\includegraphics[width=0.9\columnwidth]{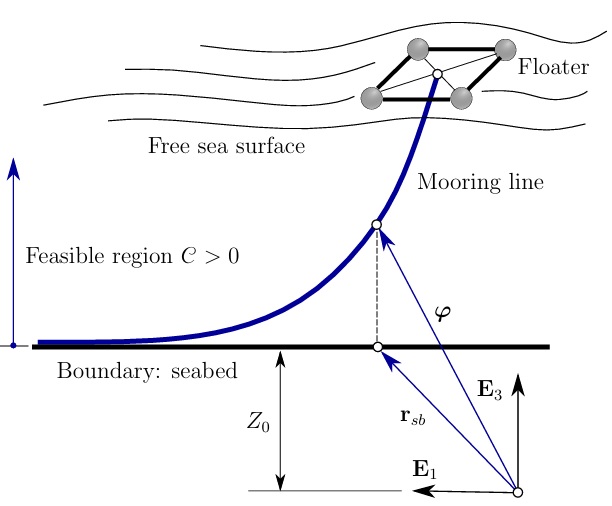}
\caption{Schematic of a mooring line attached to a floater and lying down on the seabed.}
\label{fig.barrier.1}
\end{figure}

In this subsection, we introduce a smooth unilateral constraint into the formulation of the nonlinear rod presented above via a penalty procedure. This constraint is intended for representing the ``contact'' of the mooring line with the seabed. The procedure consists of composite functions, the so-called \textit{barrier functions}, which are used to replace inequality constraints (e.g. the seabed) by introducing a penalizing term in the original functional \eqref{eq.hamilton.4}. Specifically, a barrier function is a penalty function that forces the solution to remain within the interior of the feasible region. On this basis, we modify \eqref{eq.weak.4} by introducing a penalty term as follows,
\begin{equation}\label{eq.weakmod.7}
\begin{split}
    \overline{\delta\mathcal{F}} &= \int_0^T\left[\delta\mathcal{T}(\curvep,\curved,\curvedp;t) - \delta\mathcal{U}(\curvep,\curvepp;t) \right. \\
    & -\left.\mu\,\delta\mathcal{H}(\mathcal{C}(\curve)) - \overline{\delta\mathcal{W}}\right]\,dt = 0,
    \end{split}
\end{equation}

\noindent
where $\mu$ is a positive penalty parameter, $\mathcal{C}$ is the unilateral constraint to be imposed on the rod, and $\mathcal{H}$ is a functional representing the barrier term,
\begin{equation}\label{eq.barrier.8}
    \mathcal{H}=\int_0^L\mathcal{g}(\mathcal{C})\,ds,
\end{equation}

\noindent
where $\mathcal{g}$ is a barrier function. The two most common types of barrier functions are \textit{inverse barrier functions} and \textit{logarithmic barrier functions} \cite{fiacco1990}. As we are interested in modeling mooring lines in offshore wind, it is crucial to consider the seabed as a constraint on the motion of the rod. To this end, the seabed constraint $\mathcal{C}$ can be represented mathematically by the following inequality,
\begin{equation}\label{eq.constraint.9}
    \mathcal{C}(\curve) = \left(\curve - \mathbf{r}_{sb} \right)\cdot\mathbf{E}_3 = \curve\cdot\mathbf{E}_3 - Z_0 > 0,
\end{equation}

\noindent
where $\mathbf{r}_{sb}$ is the position vector of the sea floor with respect to a global reference frame $\mathcal{E}$, and $Z_0$ is the vertical coordinate of the sea floor with respect to the global frame $\mathcal{E}$-frame (see Fig. \ref{fig.barrier.1}). Replacing \eqref{eq.constraint.9} into \eqref{eq.barrier.8} and taking the variation of the barrier term, we obtain the following expression,
\begin{equation}\label{eq.variationconst.10}
\begin{split}
   & \delta\mathcal{H}=\int_0^L\delta\mathcal{g}\left( \curve\cdot\mathbf{E}_3-Z_0\right)\,ds,\quad\text{with} \\
   &\delta\mathcal{g}(\curve\cdot\mathbf{E}_3-Z_0) = \partial_z\left.\left[\mathcal{g}(z)\right]\right\vert_{z=\mathcal{C}}\mathbf{E}_3\cdot\delta\curve,
\end{split}
\end{equation}

\subsection{Forces induced by a surrounding flow}\label{subsec:flow}

For the numerical simulation of mooring lines is also crucial account for the forces induced by the surrounding flow, namely water in rest or sea currents. To this end, we consider a simplified model consisting of fourth counterparts: the resulting force due to the accelerated fluid by the moving rod (added mass), the tangential drag force, the normal drag force, and the buoyancy force. In this regard, the resulting flow force $\mathbf{F}_f$ per unit length at an arbitrary point on the rod may be expressed as follows,
\begin{equation}\label{eq.flow.1}
\begin{split}
    \mathbf{F}_f(\curved,\curvedd;t) &= C_1\mathbf{A}_n +  C_2\abss{\mathbf{V}_n}\mathbf{V}_n + C_3\abss{\mathbf{V}_t}\mathbf{V}_t \\
    &+ C_4\left(\mathbf{V}_n + \mathbf{V}_t\right) + \mathbf{F}_{hs},
    \end{split}
\end{equation}

\noindent
where $\mathbf{F}_{hs}$ is the hydrostatic force, $\mathbf{A}_n$ is the normal component of the relative flow acceleration with respect to the rod, $\mathbf{V}_n$ and $\mathbf{V}_t$ are the normal and tangential relative flow velocity, respectively, and the coefficients $C_k$ for $k=1,\ldots,3$ are given by,
\begin{equation}\label{eq.flow.2}
\begin{split}
    C_1 = \frac{1}{4}\pi C_m \rho_{\infty} \diameter^2,\quad C_2 = \frac{1}{2} C_n \rho_{\infty} \diameter,\quad\text{and}\quad C_3 =  \frac{1}{2} C_t \rho_{\infty} \diameter,
\end{split}
\end{equation}

\noindent
with $\rho_{\infty}$ being the mass density of the surrounding flow, and $\diameter$ the diameter of the cylindrical cross-section of the rod. The coefficients $C_4$, $C_m$, $C_n$, and $C_t$ depend on the Reynolds number and are determined experimentally. 

\begin{figure}[h]
\centering
\includegraphics[width=0.95\columnwidth]{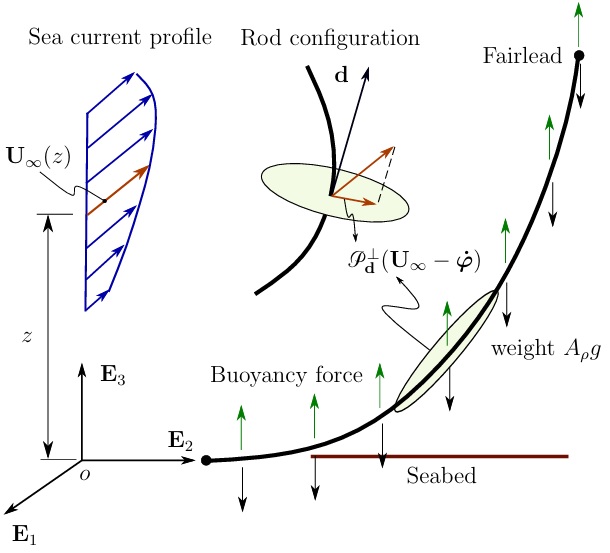}
\caption{Schematic of flow forces acting on a mooring line.}
\label{fig.flow.2}
\end{figure}

As described in \cite{Nguyen2024}, the normal and tangential components of the surrounding flowfield relative to the rod are determined by considering the kinematic description given in the previous subsection. Those quantities can be expressed as follows,
\begin{equation}\label{eq.flow.3}
\begin{split}
    \mathbf{A}_n &= \mathcal{P}_{\mathbf{d}}^{\perp} \mathbf{A}, \\
    \mathbf{V}_n &= \mathcal{P}_{\mathbf{d}}^{\perp} \mathbf{V},\quad\text{and} \\
    \mathbf{V}_t &= \mathcal{P}_{\mathbf{d}}^{\parallel} \mathbf{V},
\end{split}
\end{equation}

\noindent
where $\mathbf{V}=\mathbf{U}_{\infty}-\curved$, $\mathbf{A}=\dot{\mathbf{U}}_{\infty}-\curvedd$, and $\mathcal{P}_{\mathbf{d}}^{\parallel}=\mathbf{d}\otimes\mathbf{d}$ is the tangential projection operator in $\mathbb{R}^3$. In this work, the magnitude and direction of the sea current velocity $\mathbf{U}_{\infty}(z;t)$ is a function of the vertical coordinate $z$ and time $t$ (see Fig. \ref{fig.flow.2}). Such assumptions allow us to consider depth-dependent profiles of ocean currents. Replacing \eqref{eq.flow.3} into \eqref{eq.flow.1} and integrating along the line, we obtain the resulting flows forces as,
\begin{equation}\label{eq.flow.4}
\begin{split}
    \mathbf{F}_f &= C_1 \int_0^L\mathcal{P}_{\mathbf{d}}^{\perp}\left( \dot{\mathbf{U}}_{\infty}-\curvedd\right)\,ds \\
    &+ C_2 \int_0^L \abss{\mathcal{P}_{\mathbf{d}}^{\perp}\left(\mathbf{U}_{\infty}-\curved \right)}\mathcal{P}_{\mathbf{d}}^{\perp}\left(\mathbf{U}_{\infty}-\curved \right)\,ds \\
    &+ C_3 \int_0^L \abss{\mathcal{P}_{\mathbf{d}}^{\parallel}\left(\mathbf{U}_{\infty}-\curved \right)}\mathcal{P}_{\mathbf{d}}^{\parallel}\left(\mathbf{U}_{\infty}-\curved \right)\,ds \\
    &+ C_4\left[\int_0^L\mathcal{P}_{\mathbf{d}}^{\perp}\left(\mathbf{U}_{\infty}-\curved\right)\,ds + \int_0^L\mathcal{P}_{\mathbf{d}}^{\parallel}\left(\mathbf{U}_{\infty}-\curved\right)\,ds\right] + \mathbf{F}_{hs},
\end{split}
\end{equation}

\noindent
where the total hydrostatic force vector, $\mathbf{F}_{hs}$, is obtained by integrating the pressure all over the rod surface immersed in the flow medium. Since we are here interested in studying fully submerged mooring lines, the ``wet'' surface of the rod completely encloses its volume; therefore, the hydrostatic pressure is given by a potential field \cite{Gunnar2024}. By invoking Gauss's theorem, we obtain the following expression,
\begin{equation}\label{eq.flow.5}
\begin{split}
    \mathbf{F}_{hs} = \int_{S_w}p\mathbf{n}\,dS_w = \int_V \nabla p\,dV = \frac{1}{4}\pi\rho_{\infty}g\diameter^2 L\,\mathbf{E}_3,
\end{split}
\end{equation}

\noindent
where $p$ is the pressure acting on the surface of the rod, $g$ is the acceleration due to the gravitational field, $S_w$ is the wet area of the rod, and $\mathbf{n}$ is a unit vector normal to the rod surface. As it can be observed in \eqref{eq.flow.5}, $\mathbf{F}_{hs}$ is a pure vertical buoyancy force. In this respect, the submerged weight of the rod per unit length is simply given by $(A_{\rho}g - \rho_{\infty}gA)\mathbf{E}_3$ (see Fig. \ref{fig.flow.2}). The reader should note that mooring lines, or cables, may not be wet at one or both ends, thus preventing the volume integral in \eqref{eq.flow.5} from describing the hydrostatic pressure phenomenon. Therefore, a correction is necessary to take into account the missing pressure at the ends. Such a correction leads to the concept of \textit{effective tension}, which is well described in \cite{Gunnar2024}. However, for very long and slender structures such as mooring lines, this phenomenon is negligible and we can rule out such a correction.

In addition, it should be noticed that in fluid-structure interaction problems involving one-dimensional structural elements, such as the one used here, immersed in three-dimensional flow fields, poses a significant mathematical challenge due to the use of the trace operator to reduce the three-dimensional domain interactions to a one-dimensional framework \cite{d2008coupling,kuchta2019preconditioning}. This dimensional reduction is inherently ill-posed because information about the distribution of fluid forces along the rod's length is projected onto a lower-dimensional domain, thus leading to a potential loss of physical fidelity and numerical inconsistencies.

Despite this theoretical limitation, the application of such models in practice has proven effective, producing reasonable and reliable results for many engineering problems. The justification lies in their ability to capture the dominant physical effects with acceptable computational efficiency. 



\subsection{Spatial discretization}\label{subsec:discrete}

In this subsection, we briefly describe the spatial discretization of the rod configuration $\curve(s,t)\in\mathcal{D}$ and its variation $\delta\curve(s,t)$. For this purpose, we utilize an approach based on the well-known isogeometric analysis, which the higher-order continuity of smooth spline functions allows lower polynomial degree and omit the director as a field variable. Along this path, $\curve$ is approximated by a weighted finite sum of $n_{cp}$ B-splines $N_k(s)\in C^r$ of polynomial degree $p$, where $1\leq r \leq p-1$, i.e.,
\begin{equation}\label{eq.spdis.1}
\begin{split}
    \curve(s,t) &\approx\curve_h(s,t)=\sum_k^{n_{cp}}N_k(s)\mathbf{x}_k(t)=\mathbf{N}\mathbf{q},\quad\text{and} \\
    \delta\curve(s,t) &\approx \delta\curve_h(s,t) = \sum_k^{n_{cp}}N_k(s)\delta\mathbf{x}_k(t)=\mathbf{N}\delta\mathbf{q},
\end{split}
\end{equation}

\noindent
where $\curve_h$ denotes the discrete rod configuration in the three-dimensional vector space $\mathbb{R}^3$, $\mathbf{x}_k\in\mathbb{R}^3$ is the time-dependent position vector of the $k$-th control point, $\mathbf{q}=\mathbf{q}(t)\in\mathbb{R}^{3n_{cp}}$ is the global vector of unknown time-dependent coefficients, $\mathbf{N}(s)\in\mathbb{R}^{3\times 3n_{cp}}$ is the global shape function matrix, and $\delta\curve_h$, $\delta\mathbf{x}_k$, and $\delta\mathbf{q}$ are the variations of the quantities introduced above. Replacing \eqref{eq.spdis.1} into the variational formulation \eqref{eq.EoMweak} and considering \eqref{eq.variationconst.10}, we obtain the following semi-discrete formulation,
\begin{equation}\label{eq.spdis.2}
\begin{split}
    &\int_0^L \delta\mathbf{q}\cdot\left[ \mathbf{M}(\mathbf{q})\nabla_{\dot{\mathbf{q}}}\dot{\mathbf{q}} + \mathbf{B}^T(\mathbf{q})\boldsymbol{\sigma}_h - \mu\mathbf{H}(\mathbf{q})\mathbf{E}_3 -\right.\\
   & - \left.\mathbf{N}^T(s)\mathbf{f}_{ext} \right]\,ds = 0, \quad\forall\delta\mathbf{q}\in\mathbb{R}^{3n_{cp}},
\end{split}
\end{equation}

\noindent
where $\mathbf{H}(\mathbf{q})=\mathbf{N}^T(s)\partial_z\left.\left[g(z)\right]\right\vert_{z=\mathcal{C}(\mathbf{\curve})}\in\mathbb{R}^{3n_{cp}\times 3}$ is the \textit{penalization matrix}, which represents the term associated with the barrier function. $\mathbf{M}$ and $\mathbf{B}$ are the mass matrix and the discrete version of the strain operator, respectively, whose expressions can be found in \cite{Nguyen2024}. 


\subsection{Time integration scheme}\label{subsec:timescheme}

In this work, we use the same implicit second-order numerical scheme as in \cite{gebhardt2021,Nguyen2024}, which is a hybrid combination of the midpoint and trapezoidal rules. In \cite{gebhardt2020}, it was showed that such a scheme exactly preserves the linear and angular momenta and approximately preserves energy. To conduct dynamic simulations, we evaluate the semi-discrete set of ODEs \eqref{eq.spdis.2} at time instant $t_{n+\frac{1}{2}}\in[t_n,t_{n+1}]$ as follows,
\begin{equation}\label{eq.time.1}
\begin{split}
    &\int_0^L \left( \left[\mathbf{M}(\mathbf{q})\nabla_{\dot{\mathbf{q}}}\dot{\mathbf{q}}\right]_{n+\frac{1}{2}} + \left[\mathbf{B}^T(\mathbf{q})\boldsymbol{\sigma}_h\right]_{n+\frac{1}{2}}\right. \\ 
    & - \left.\left[\mu\mathbf{H}(\mathbf{q})\mathbf{E}_3\right]_{n+\frac{1}{2}} - \left[\mathbf{N}^T(s)\mathbf{f}_{ext}\right]_{n+\frac{1}{2}} \right)\,ds = \mathbf{0}.
\end{split}
\end{equation}

In the previous expression, the inertial term $\mathbf{M}(\mathbf{q})\nabla_{\dot{\mathbf{q}}}\dot{\mathbf{q}}$ evaluated at $t_{n+\frac{1}{2}}$ is approximated by using an extended version of the midpoint rule as,
\begin{equation}\label{eq.time.2}
\begin{split}
   \left[\mathbf{M}(\mathbf{q})\nabla_{\dot{\mathbf{q}}}\dot{\mathbf{q}}\right]_{n+\frac{1}{2}} &\approx \frac{\mathbf{M}_{n+1}\dot{\mathbf{q}}_{n+1} - \mathbf{M}_n\dot{\mathbf{q}}_n}{\Delta t} \\ 
   & -\frac{1}{2}\partial_{\boldsymbol{\eta}}\left( \dot{\mathbf{q}}_{n+\frac{1}{2}}\cdot\mathbf{M}(\boldsymbol{\eta})\cdot\dot{\mathbf{q}}_{n+\frac{1}{2}}\right)_{\boldsymbol{\eta}=\mathbf{q}_{n+\frac{1}{2}}},
\end{split}
\end{equation}

\noindent
along with the classical approximation formulas,
\begin{equation}\label{eq.time.3}
\begin{split}
   \mathbf{q}_{n+\frac{1}{2}} \approx \frac{\mathbf{q}_n + \mathbf{q}_{n+1}}{2},\quad\text{and}\quad \dot{\mathbf{q}}_{n+\frac{1}{2}} \approx \frac{\mathbf{q}_{n+1} - \mathbf{q}_n}{\Delta t},
\end{split}
\end{equation}

\noindent
where $\Delta t$ is the time step. The internal forces $\mathbf{B}^T(\mathbf{q})\boldsymbol{\sigma}$ and penalization term $\mu\mathbf{H}(\mathbf{q})\mathbf{E}_3$ at $t_{n+\frac{1}{2}}$ are approximated by the trapezoidal rule as,
\begin{equation}\label{eq.time.4}
\begin{split}
   \left[\mathbf{B}^T(\mathbf{q})\boldsymbol{\sigma}\right]_{n+\frac{1}{2}} &\approx \frac{1}{2}\left(\mathbf{B}_n^T\boldsymbol{\sigma}_n + \mathbf{B}_{n+1}^T\boldsymbol{\sigma}_{n+1} \right), \\
   \left[ \mu\mathbf{H}(\mathbf{q})\mathbf{E}_3\right]_{n+\frac{1}{2}} &\approx \frac{\mu}{2}\left( \mathbf{H}_n + \mathbf{H}_{n+1}\right) \mathbf{E}_3.
\end{split}
\end{equation}

Once $\mathbf{q}_{n+1}$ has been determined through an iterative method, for example, the Newton-Raphson method, we compute $\dot{\mathbf{q}}_{n+1}$ as $\frac{2}{\Delta t}(\mathbf{q}_{n+1}-\mathbf{q}_n)-\dot{\mathbf{q}}_n$. As mentioned before, the implicit nature of the numerical integration scheme presented above requires the utilization of iterative procedures to solve the time-discrete set of equations \eqref{eq.time.1}. Furthermore, iterative schemes such as the Newton-Raphson method demands each term in \eqref{eq.time.1} to be linearized. The reader can find the full expressions for the tangent stiffness matrices associated with the inertial term, internal forces term and configuration-dependent forces (including those coming from the surrounding flow) in \cite{Nguyen2024}. The tangent stiffness matrix associated to the penalization term can be obtained by following the same procedure as in \cite{Nguyen2024}.


\section{Numerical results}\label{resultsSec}

In this section, we present a series of results regarding the behavior of the advanced nonlinear Kirchhoff rod model described above when studying mooring lines for the offshore wind energy sector (hereinafter referred to as ARMoor). First, we present a full comparison against well-established problems in the literature, such as the elastic catenary under gravity loads. In addition, we test our ``contact model'' by comparing the touchdown point (between the mooring line and the seabed) and reaction forces at the fairlead against solutions reported in \cite{Gunnar2024}. Second, we consider a mooring line under the action of a pulsating force acting along and normal to the tangent vector to the rod centerline at the fairlead. Finally, we consider the IEA 15MW offshore wind turbine on the UMaine VolturnUS-S marine platform \cite{allen2020}. We then conduct dynamic simulations of the mooring line system by using our nonlinear rod model and we assess its performance when compared  against numerical results obtained with OpenFAST.  



\subsection{Case 1: elastic catenary comparison}\label{subsec:cat}

In this subsection, we consider two different cases: \textit{i}) a hanging rod between two supports at a different level (hereinafter referred to as ROD-A), and \textit{ii}) a mooring line resting on the seabed, subject to conservative and non-conservative forces (gravitational and buoyancy loads + fluid forces) where the fairlead position/force is specified (hereinafter referred to as ROD-B). It should be noted that case ROD-A consists only of a classical static analysis, However, for case ROD-B, we find the static equilibrium of the rod by letting it to relax dynamically to a static position. 

For case ROD-A, we consider a cable DuPont's Kevlar 49 type 968, whose axial elastic modulus is $E=81.8$ GPa \cite{roccia2020cable}. The rest of the parameter are density $A_{\rho}=0.055$ kg/m, cable length $L=300$ m, cross-section diameter $\diameter = 0.007$ m, gravity constant $g=9.81$ m/s$^2$, and boundary conditions $\curve(0)=(0,0,0)^T$ m and $\curve(L)=(100,0,50)^T$ m. The objective of this showcase is to test the performance of our rod model, when modeling very flexible cables, against the three-dimensional elastic catenary solution given in \cite{impollonia2011}.

\begin{figure}[H]
\centering
\includegraphics[width=0.99\columnwidth]{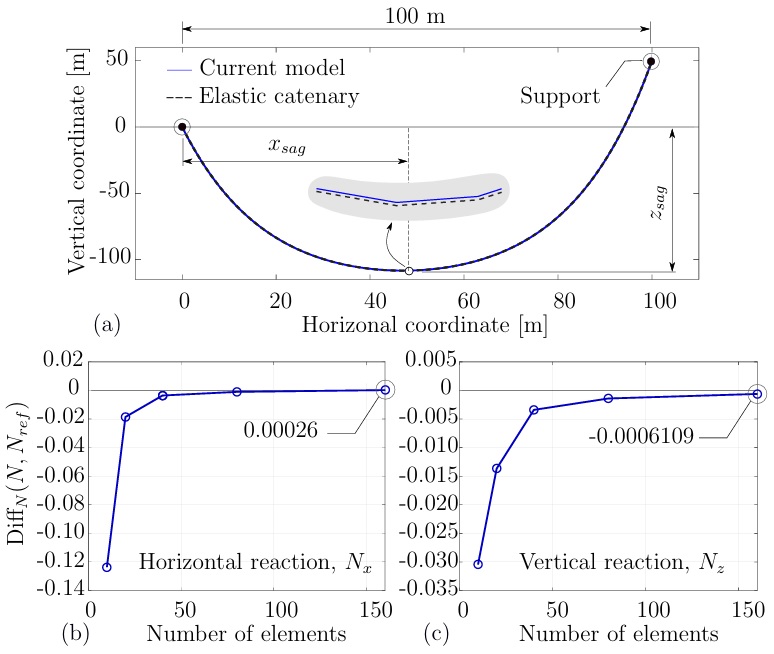}
\caption{(Top) Comparison of the final configuration between the three-dimensional elastic catenary and our current model. (Bottom) convergence analysis for the horizontal and vertical reaction forces.}
\label{fig.cat.1}
\end{figure}

In Fig. \ref{fig.cat.1}a, we present the final configuration for a cable hanging between two supports obtained by using our nonlinear rod model and the well-known three-dimensional elastic catenary model \cite{impollonia2011}. For this example, we assumed a spatial discretization of $N_e=40$ elements and the following settings for the solver: polynomial degree $p=3$, continuity $r=1$, tolerance error for Newton iterations $e_{tol}=10^{-10}$, and number of step loads $N_{\Delta F}=500$. Although the models are certainly different, they show an excellent agreement from a geometric shape point of view. However, as we zoom in, a slight difference can be observed between the two solutions. Specifically, the numerical solution obtained by ARMoor is bounded from below by the one obtained with the elastic catenary. This behavior is expected since our rod model considers bending stiffness, while the catenary does not. As consequence, our solution is stiffer exhibiting a lower elongation in the deformed configuration (see Table \ref{Catenary.T1}). It should be mentioned that if the bending contribution is neglected, our model reduces to the well-known equations of the elastic catenary. 

Finally, in Fig. \ref{fig.cat.1}b and \ref{fig.cat.1}c, we present a simple convergence analysis of the reaction forces at the upper right support, i.e. $\curve(L)=(100,0,50)^T$ m. To this end, we define the following criteria to evaluate the difference between solutions,
\begin{equation}\label{eq.diff.1}
\begin{split}
\text{Diff}_{X}(X,X_{ref})&:=\frac{X-X_{ref}}{X_{ref}} = \frac{X}{X_{ref}} - 1, \\
   \overline{\text{Diff}}_{X}(X,X_{ref})&:=\left\vert\text{Diff}_{X}(X,X_{ref})\right\vert\cdot 100\%, \\
   \text{MAD}_X(X,X_{ref})&:=\frac{1}{n_p}\sum_{i=1}^{n_p}\left\vert  X_i - X_{ref,i}\right \vert,  
\end{split}
\end{equation}

\noindent
where $\text{Diff}_{X}$ represents the relative difference between the prediction $X$ with respect to the reference value $X_{ref}$, $\overline{\text{Diff}}_{X}$ its percentage difference in absolute value, and $\text{MAD}_X$ stands for mean absolute difference. This quantity is similar to the mean absolute error (MAE) which, in statistics, represents a measure of errors between paired observations expressing the same phenomenon. As it can be observed in Fig. \ref{fig.cat.1}, the horizontal and vertical reactions forces at $s=L$ obtained through our rod model approaches the classical elastic catenary solution as the number of elements increases. Specifically, for $N_e=160$, the difference for $N_x$ and $N_z$ is $0.026\%$ and $0.061\%$, respectively. Table \ref{Catenary.T1} lists the values for the sag position, rod elongation and reaction forces for $N_e=40$. As can be observed, all of them are in excellent agreement with those values computed using the elastic catenary model.

\begin{table*}[h!]
\centering
\caption{IGA solution vs classical elastic catenary.}
\label{Catenary.T1}
\begin{tabular}{ >{\arraybackslash}m{.25\linewidth} 
>{\centering\arraybackslash}m{.16\linewidth} 
>{\centering\arraybackslash}m{.16\linewidth} >{\centering\arraybackslash}m{.14\linewidth}}
\hline\hline
Solution & Catenary & ARMoor ($N_e=40$)  & $\overline{\text{Diff.}}_X(\cdot,\cdot)$ $[\%]$\\ 
\hline
Horiz. sag, $x_{sag}$ [m] & 46.69259 & 46.694272 & 0.00359  \\ [0.1cm]
Vert. sag, $z_{sag}$ [m] & -108.3255 & -108.305688 & 0.01836 \\[0.1cm]
Elongation, $\Delta L$ [m] & 0.00414399 & 0.00414377 & 0.00545\\[0.1cm]
Horiz. reaction, $N_x$ [N] & 9.576918 & 9.543122 & 0.35288 \\[0.1cm]
Vert. reaction, $N_z(L)$ [N] & 94.51768 & 94.196588 & 0.48396\\[0.1cm]
\hline\hline
\end{tabular}
\end{table*}

\begin{figure}[H]
\centering
\includegraphics[width=0.99\columnwidth]{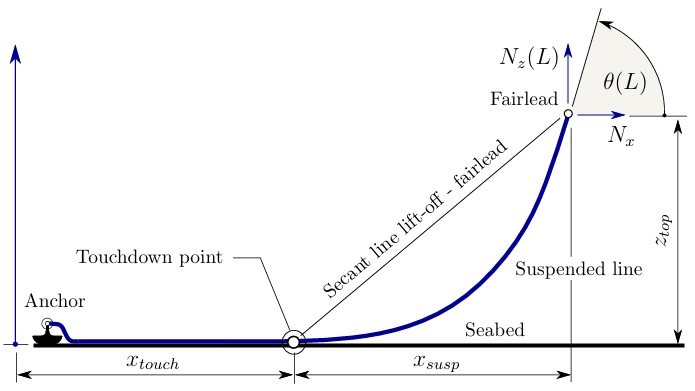}
\caption{Schematic of a mooring line lying down on the seabed.}
\label{fig.mooring.1}
\end{figure}

For case ROD-B, we consider the mooring line described in \cite[Ch. \ 7.6.4, p. 257]{Gunnar2024}, which is characterized by a line length $L=627$ m, submerged weight per unit length $w_s=2.46$ kN/m, and axial stiffness $EA=892.6$ MN. For the fluid surrounding the mooring line, we assume: water density $\rho_{\infty}=1000$ kg/m$^3$, normal drag coefficient $C_n=1.2$, tangential drag coefficient $C_t=0.062$, added mass coefficient $C_m=1.0$, linear normal drag coefficient $C_4=C_n$, and free sea current velocity and acceleration $\mathbf{U}_{\infty}=\dot{\mathbf{U}}_{\infty}=\mathbf{0}$ (fluid in rest).  In addition, the vertical position of the fairlead is located at $z_{top}=71.2$ m above the seabed. Fig. \ref{fig.mooring.1} presents a schematic of a mooring line placed on the seabed where the main parameters to be used for comparison are highlighted.

As mentioned above, for this showcase, we conduct a dynamic simulation to find out the ``final static'' configuration of the mooring line. Typically, this sort of study requires a longer simulation time to allow the rod to relax (by dissipating energy through the surrounding flow) to an equilibrium position. As it is well known, the barrier function introduced into the functional \eqref{eq.weakmod.7} modifies the original ``\textit{energy landscape}'' of \eqref{eq.weak.4} by adding a new component that represents a penalty. While this additional energy (or cost) discourages the solution from violating constraints, it introduces an ``artificial'' potential to enforce which here is called \textit{seabed constraint}. As consequence, the static problem,
\begin{equation}\label{eq.static.1}
    \int_0^L\delta\curve\cdot[\mathcal{B}^T(\curvep, \curvepp)\boldsymbol{\sigma}-\mu\mathcal{G}(\curve)-\mathbf{f}_{ext}]ds=0,
\end{equation}

\noindent
with $\mathcal{G}(\curve)=\partial_z[\mathcal{g}(z)]\vert_{z=\mathcal{C}}\mathbf{E}_3$, becomes significantly unstable from a numerical point of view. Such phenomenon could be explained by either large gradients of the barrier function near the constraint boundary or conditioning problems associated with the Hessian of the objective function. Consequently, static equilibrium positions are quite challenge to obtain by solving \eqref{eq.static.1} (non-convergence of the Newton method), reason why we adopt a dynamic relaxation approach, where the mass and dissipation from the fluid act as stabilizing terms which allows solving the penalized problem.

Among all the possible barrier functions in the context of computational mechanics and optimization, we can mention: the logarithmic, reciprocal, quadratic, exponential, and hyperbolic barrier functions along with some combinations among them such as the log-quadratic penalty function (see Table \ref{Barrier.T2}) \cite{fiacco1990,wriggers2006}.  

\begin{table}[h!]
\centering
\caption{Barrier functions.}
\label{Barrier.T2}
\begin{tabular}{ >{\arraybackslash}m{.19\linewidth} 
>{\centering\arraybackslash}m{.24\linewidth} 
>{\raggedright\arraybackslash}m{.47\linewidth}}
\hline\hline
Function & Expression & Description \\ 
\hline
Logarithmic & $-\log(f(z))$ & {\small The function is effective in ensuring that $g(f(z))>0$, as it becomes infinitely large when $f(z)\longrightarrow0$} \\ [0.1cm]
\hline
Reciprocal & $\frac{1}{f(z)}$ & {\small Like the logarithmic barrier, the reciprocal barrier function penalizes solutions as $f(z)\longrightarrow0$} \\[0.1cm]
\hline
Quadratic & $\alpha f(z)^2$ & {\small Common in penalty methods where $\alpha>0$ is a parameter that controls the ``strength'' of the barrier} \\[0.1cm]
\hline
Exponential & $e^{-f(z)}$ & {\small This barrier function is used to create a smooth transition in penalties. It becomes very steep as $f(z)$ approaches zero} \\[0.1cm]
\hline
Hyperbolic & $\frac{1}{f(z)^q}$ for $q>1$ & {\small The hyperbolic barrier is a generalization that includes steepness control through the parameter $q$} \\[0.1cm]
\hline
Log-quadratic & $\log(f(z))+\frac{\alpha}{2} f(z)^2$ & {\small This is a hybrid approach combining both logarithmic and quadratic penalties to balance soft and hard constraints} \\[0.1cm]
\hline\hline
\end{tabular}
\end{table}

Based on the empirical investigation carried out in \cite{Veseth2023}, we first explore the behavior of the logarithmic barrier function and the reciprocal barrier function to penalize the solution when the rod approaches the seabed. To this end, we conduct a parametric study for $\mu$ ranging from $0.5$ to $1000$ considering a spatial discretization $N_e=128$ and force/displacement steps $N_{\Delta F}=N_{\Delta u}=500$. Besides understanding the behavior of these barrier functions for different values of $\mu$, the main motivation behind this analysis is to properly calibrate $\mu$ for case ROD-B outlined above. Regarding the solver configuration, we assume the same settings as for ROD-A. 

\begin{figure}[h!]
\centering
\includegraphics[width=0.99\columnwidth]{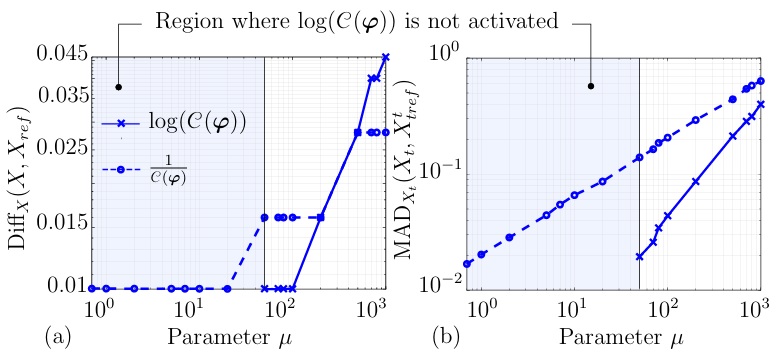}
\caption{Force control study. (a) Difference for the touchdown as a function of penalty parameter $\mu$ for two different barrier functions when compared to the reference solution \cite{Gunnar2024}. (b) Mean absolute difference between the portion of the cable in ``contact'' with the seabed and the vertical coordinate of the seabed $Z_0$ (gap) as a function of $\mu$. }
\label{fig.parameterMu.1}
\end{figure}

To bring the fairlead to its final position, we use both a force control (FC) and displacement control (DC) strategy. The quality of our formulation when considering the barrier term is then assessed by comparing our solutions against results obtained using a catenary-based approach well suited for mooring line applications \cite{Gunnar2024}. In Fig. \ref{fig.parameterMu.1}a, we present the relative difference $\text{Diff}_X$ between our model and the reference one for the touchdown point (TP) when using a force control strategy. Here, we can observe a region, $\mu<50$, where the barrier function $\log(\mathcal{C}(\curve))$ is not activated, i.e. the term $\mu\mathbf{H}(\mathbf{q})\mathbf{E}_3$ might not significantly impact the overall objective function. As a result, the constraint is violated because the penalty is not large enough to prevent the system from crossing the boundary. For values of $\mu$ bigger than $50$, the logarithmic barrier function performs similarly to the reciprocal function. In Fig. \ref{fig.parameterMu.1}b, we plot the mean absolute difference $\text{MAD}_X$ between the portion of the cable lying down on the seabed and the seafloor itself (coordinate $Z_0$) as function of $\mu$. As expected, we observe that for $\mu<50$ the logarithmic barrier function does not activate, while the alternative $1/\mathcal{C}(\curve)$ behaves linearly as a function of $\mu$.  

\begin{figure}[h!]
\centering
\includegraphics[width=0.99\columnwidth]{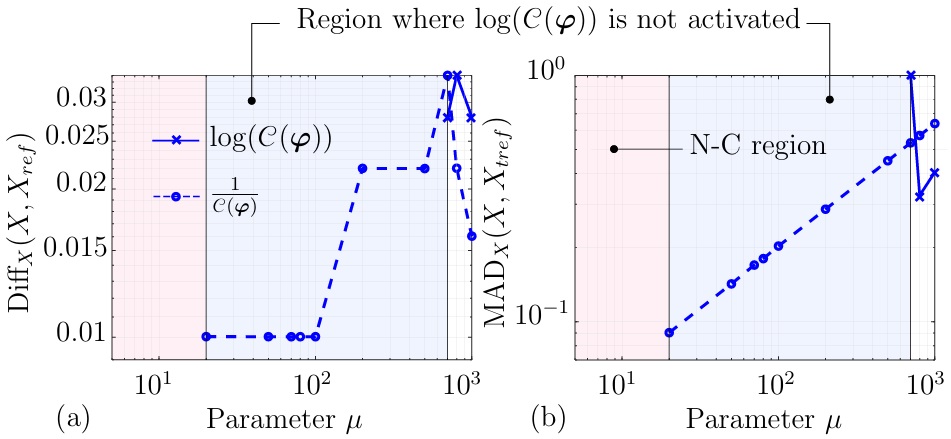}
\caption{Displacement control study. (a) Difference for the touchdown as a function of penalty parameter $\mu$ for two different barrier functions when compared to the reference solution \cite{Gunnar2024}. (b) Mean absolute difference between the portion of the cable in ``contact'' with the seabed and the vertical coordinate of the seabed $Z_0$ (gap) as a function of $\mu$. }
\label{fig.parameterMu.2}
\end{figure}

In Fig. \ref{fig.parameterMu.2}, we present the same study as before, but now considering a displacement control approach to move the fairlead to its final position. Here, we observe that for penalty parameters lower than $20$, the Newton-Raphson scheme does not converge when using the reciprocal barrier function (N-C region). Furthermore, we notice that the logarithmic barrier function is only activated for $\mu>700$, i.e. more than one order of magnitude compared to the force control strategy. This unstable behavior for $\mu<20$ for $1/\mathcal{C}(\curve)$ and $\mu<700$ for $\log(\mathcal{C}(\curve))$ when using displacement control may be associated with very large displacement increments, thus requiring a significant contribution from the penalty term at the beginning of the simulation to prevent the line from violating the constraint. In this sense, small values of $\mu$ may not be sufficient to activate the barrier function, or even worse, make the numerical scheme unstable. Along this path, we conducted a numerical experiment to examine the effect of reducing the displacement increment on the activation of the barrier function. Our findings show that as the displacement increment decreases, the value of the parameter $\mu$ required to activate the barrier function also decreases. For instance, when using a number of displacement steps of $N_{\Delta u}=2000$, the value of $\mu$ decreases from 700 to 500. This suggests that larger displacement steps lead the barrier function to take very small values, thus requiring a higher penalty parameter to prevent the line from violating the constraint. 

Based on the previous analysis, for ROD-B, we select the reciprocal barrier function since it behaves better, for lower values of $\mu$, than the logarithmic barrier function. Furthermore, we adopt $\mu=25$ and a total simulation time of $T=80$ s, which is divided into two intervals as $T=T_1+T_2$. During the first interval, $T_1=30$ s, the fairlead is brought into its final position, while during the second simulation interval, $T_2 = 50$ s, the solution is allowed to dynamically relax to its static equilibrium position.

Here, we consider six different cases by specifying the final horizontal force or final fairlead position according to the strategy used, force control or displacement control, respectively. In Table \ref{Cases.T3}, we describe such cases discriminated according to whether FC or DC is used. It is worth mentioning that for DC, the vertical position of the fairlead is $\curve(L)=71.2$ m for all cases.

\begin{table}[h!]
\centering
\caption{ROD-B cases.}
\label{Cases.T3}
\begin{tabular}{ >{\arraybackslash}m{.15\linewidth} 
>{\centering\arraybackslash}m{.15\linewidth}
>{\centering\arraybackslash}m{.15\linewidth}
>{\centering\arraybackslash}m{.25\linewidth}}
\hline\hline
\multirow{2}{*}{Case} & \multicolumn{2}{c}{FC [kN]} & \multirow{2}{*}{DC [m] ($\curve_x(L)$)}\\
\cline{2-3}
\multicolumn{1}{l}{} & $N_x(L)$ & $N_z(L)$ & \\
\hline
Case 1 & 100 & 256.314 & 590.781 \\ [0.1cm]
Case 2 & 1040.404 & 628.167 & 614.317 \\[0.1cm]
Case 3 & 2030.303 & 860.274 & 618.727 \\[0.1cm]
Case 4 & 3020.202 & 1041.527 & 621.149 \\[0.1cm]
Case 5 & 4010.101 & 1195.295 & 622.895 \\[0.1cm]
Case 6 & 5000 & 1331.136 & 624.316 \\[0.1cm]
\hline\hline
\end{tabular}
\end{table}

\begin{figure}[h!]
\centering
\includegraphics[width=0.99\columnwidth]{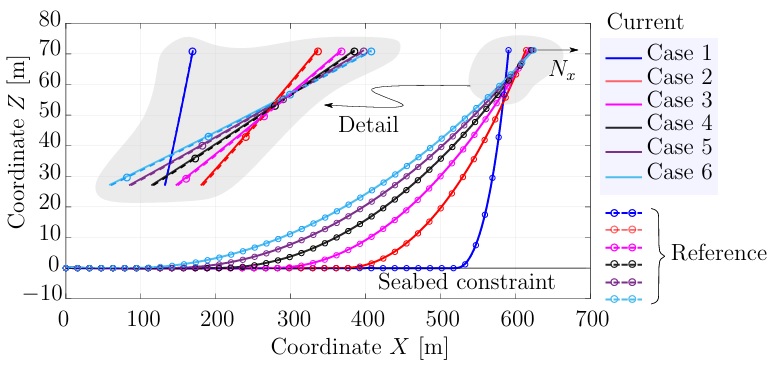}
\caption{Final configuration for the six mooring line cases outlined in Table \ref{Cases.T3} (displacement control).}
\label{fig.config.3}
\end{figure}

In Fig. \ref{fig.config.3}, we present the final line configurations for cases 1 to 6 using displacement control. As shown in this figure, our numerical solution shows excellent agreement with the reference solution from \cite{Gunnar2024}. For case 1, the maximum difference between models under DC appears in the vertical component of tension at the fairlead (approximately $2\%$), whereas under FC, the maximum difference occurs at the touchdown point, with a difference around $1.0\%$ (see Tables \ref{Cases.T4} and \ref{Cases.T5}). For the remaining cases, the differences between models remain minimal, with a maximum discrepancy of approximately $4\%$ observed at the touchdown point in case 3, regardless of whether displacement control or force control is applied. It is worth noting that all values reported in Tables \ref{Cases.T4} and \ref{Cases.T5} were averaged over the last 200 time steps of the simulation to mitigate the impact of local peaks caused by solution oscillations. This phenomenon arises as a direct consequence of the dynamic relaxation strategy used to approximate the static equilibrium position of the mooring line. Although the surrounding flow acts as a ``damper'' on the rod by dissipating energy, it is insufficient to eliminate all oscillations within the simulation time windows utilized. For comparison purposes, the results for cases 3 and 6 are presented in Tables \ref{Case3.TA1} to \ref{Case6.TA4} in \ref{appeA} for both DC and FC. 

\begin{table}[h!]
\centering
\caption{ROD-B (case 1) - Displacement control.}
\label{Cases.T4}
\begin{tabular}{ >{\arraybackslash}m{.35\linewidth} 
>{\centering\arraybackslash}m{.14\linewidth} 
>{\centering\arraybackslash}m{.18\linewidth} >{\centering\arraybackslash}m{.14\linewidth}}
\hline\hline
Solution & Ref. sol. \cite{Gunnar2024}  & ARMoor ($N_e=128$)  & $\overline{\text{Diff.}}_X(\cdot,\cdot)$ $[\%]$ \\ 
\hline
$x_{touch}$ [m] & 522.55 & 517.3176 & 1.001  \\ [0.1cm]
$\theta(L)$ [$^{\circ}$] & 68.687 & 68.217 & 0.684 \\[0.1cm]
Stretched length, $L_f$ [m] & 627.08 & 627.075 & 7.97$\times 10^{-4}$ \\[0.1cm]
$N_x(L)$ [kN] & 100 & 98.939 & 1.0610 \\[0.1cm]
$N_z(L)$ [kN] & 256.3144 & 251.398 & 1.918 \\[0.1cm]
\hline\hline
\end{tabular}
\end{table}

\begin{table}[h!]
\centering
\caption{ROD-B (case 1) - Force control.}
\label{Cases.T5}
\begin{tabular}{ >{\arraybackslash}m{.35\linewidth} 
>{\centering\arraybackslash}m{.14\linewidth} 
>{\centering\arraybackslash}m{.18\linewidth} >{\centering\arraybackslash}m{.14\linewidth}}
\hline\hline
Solution & Ref. sol. \cite{Gunnar2024}  & ARMoor ($N_e=128$)  & $\overline{\text{Diff.}}_X(\cdot,\cdot)$ [\%] \\ 
\hline
$x_{touch}$ [m] & 522.55 & 517.3244 & 1.00  \\ [0.1cm]
$\theta(L)$ [$^{\circ}$] & 68.687 & 68.288 & 0.58 \\[0.1cm]
Stretched length, $L_f$ [m] & 627.08 & 627.081 & 1.59$\times 10^{-4}$ \\[0.1cm]
$x_{touch}+x_{susp}$ [m] & 590.781 & 590.9463 & 0.028 \\[0.1cm]
$z_{top}$ [m] & 71.2 & 70.7802 & 0.589 \\[0.1cm]
\hline\hline
\end{tabular}
\end{table}

To close this subsection, we emphasize the substantial shifts in touchdown positions resulting from small changes in the fairlead point position. By shifting the horizontal fairlead position from $590.781$ m to $624.316$ m, the horizontal tension at this point increases by approximately $4900$ kN, while the corresponding change in the touchdown position is about $432.20$ m. Furthermore, the largest elastic elongation of the line is only $0.56\%$ when considering case 6, i.e. the line length increases from $627$ m to $630.54$ m. Such minimal elongations, despite large force-displacement responses, indicate that mooring line stiffness is governed more by changes in geometry than by stretching. Although these results are derived from a DC analysis, similar behavior is observed with an FC-based approach. 


\subsection{Case 2: pulsating forces}\label{subsec:case2}

In this subsection, we consider a problem of theoretical interest that also holds potential relevance in offshore wind engineering, particularly when analyzing the dynamic behavior of mooring lines connected to floating structures. Specifically, we investigate the response of a long mooring line subjected to the combined effects of gravitational forces, the surrounding fluid (water at rest), seabed contact (modeled through the previously discussed barrier term), and a pulsating force applied at the fairlead. Mooring lines serve as critical load-transferring elements between floating structures (e.g. offshore platforms and floating wind turbines) and the seabed. In this regard, pulsating forces at the fairlead often arise from platform motions or external wave and wind loads. Understanding how these forces influence mooring line dynamics is crucial for several aspects, including structural integrity, platform stability analysis, mitigation of undesirable nonlinear phenomena, and the development of effective control and monitoring strategies, among others.

\begin{figure*}[h!]
\centering
\includegraphics[width=0.79\textwidth]{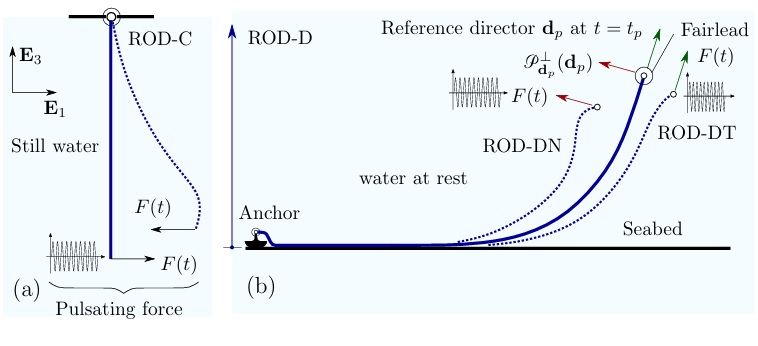}
\caption{(a) ROD-C case: schematic of a hanging line configuration. (b) ROD-D case: schematic of a mooring line under a pulsating force applied at the fairlead.}
\label{fig.pulsating.1}
\end{figure*}

Here, we conduct two different studies. First, we analyze the behavior of a simple swinging line immersed in water at rest, exposed to gravitational forces and a horizontal harmonic force with varying frequency applied at the lower end (hereinafter referred to as ROD-C; see Fig. \ref{fig.pulsating.1}a). This case is designed to verify the numerical implementation of ARMoor under harmonic loading by ensuring that the change in the system's total energy $\Delta\mathcal{E}$ through a time step $\Delta t = t_{n+1}-t_n$ matches the work $\mathcal{W}_{\Delta t}^{nc}$ done by the external forces throughout the simulation, i.e., $\Delta\mathcal{E} 
= \mathcal{W}_{\Delta t}^{nc}$. Second, we examine a mooring line in still water subjected to a pulsating force applied at the fairlead (hereinafter referred to as ROD-D). For this case, we consider two scenarios: \textit{i}) a driving force acting along the initial tangential direction to the rod's centerline at the fairlead (denoted as ROD-DT; see Fig. \ref{fig.pulsating.1}b), and \textit{ii}) the same system subjected to a driving force acting normal to the initial director of the rod at the fairlead (denoted as ROD-DN; see Fig. \ref{fig.pulsating.1}b).

\begin{figure}[h!]
\centering
\includegraphics[width=0.99\columnwidth]{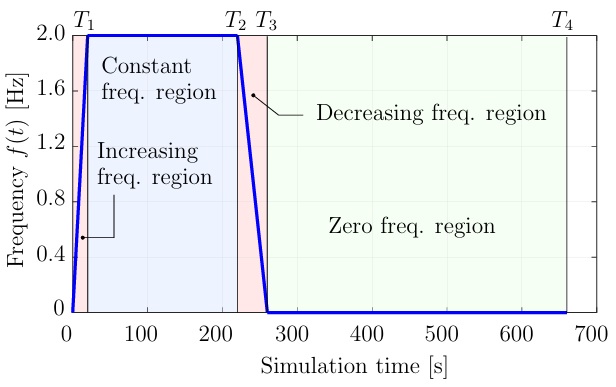}
\caption{Frequency of the driving force over time.}
\label{fig.pulsating.2}
\end{figure}

For ROD-C, we consider a rod with a length of $L=250$ m, a diameter of $\diameter=0.02$ m, a mass density per unit length of  $A_{\rho}=0.8482$ kg/m, an axial stiffness of $EA=21.99$ MN, and a bending stiffness of $EI=549.77$ Nm$^2$. The parameters associated with the surrounding water are the same as those described in the previous subsection. The director in the reference configuration is defined as $\mathbf{D}=(0,0,-1)^T$, which is aligned with the global basis vector $\mathbf{E}_3$ (see Fig. \ref{fig.pulsating.1}a). A pulsating force $\mathbf{f}(t)$ is applied at the free end along the basis vector $\mathbf{E}_1$ and is defined as follows:
\begin{equation}\label{eq.pulsating.1}
    \mathbf{f}(t) = F(t)\mathbf{E}_1 = A_F \sin\left( 2\pi f(t) t\right) \mathbf{E}_1,
\end{equation}

\noindent
where $A_F$ is the force amplitude and $\omega(t)$ is the force frequency considered to vary over time. Here we assume a constant force amplitude $A_F=175$ kN while the frequency is defined by the following piecewise function, 
\begin{equation}\label{eq.pulsating.2}
    f(t) = 
    \begin{cases} 
     f_{max} \frac{t}{T_1}, & \text{if}\,\, t\in[0,T_1] \\
     f_{max}, & \text{if}\,\, t\in(T_1,T_2] \\
     \frac{f_{max}}{T_3-T_2}\left( T_3 - t\right), & \text{if}\,\, t\in(T_2,T_3] \\
     0, & \text{if}\,\, t\in(T_3,T_4],
    \end{cases}
\end{equation}

\noindent
where $f_{max}=2$ Hz, $T_1=20$ s, $T_2=220$ s, $T_3=260$ s, and $T_4=660$ s. In Fig. \ref{fig.pulsating.2}, we plot the piecewise function \eqref{eq.pulsating.2} to illustrate how the frequency of the pulsating driving force $f(t)$ varies over time. 

In this case, we assumed a spatial discretization of $N_e=20$ elements and the following settings for the solver: polynomial degree $p=3$, continuity $r=1$, time step $\Delta t=0.0025$ s, and tolerance error for Newton iterations $e_{tol}=10^{-10}$. 

\begin{figure}[h!]
\centering
\includegraphics[width=0.99\columnwidth]{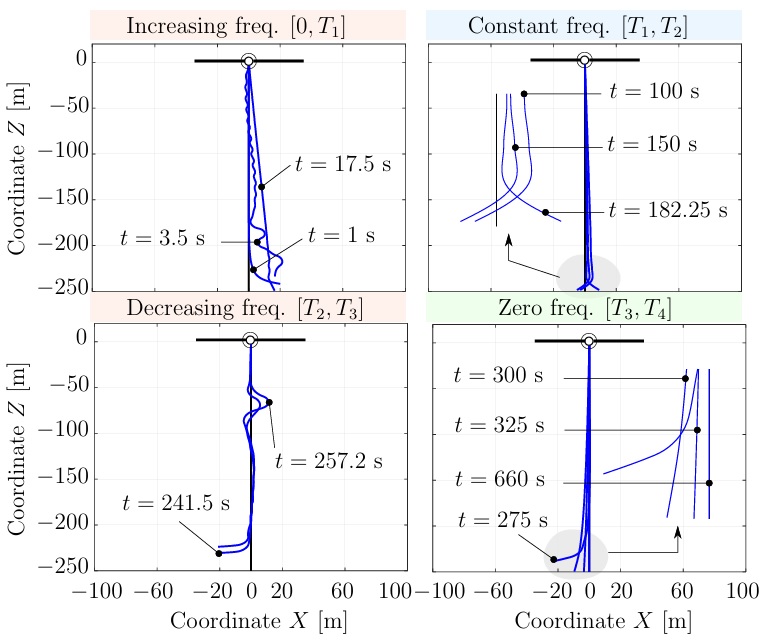}
\caption{Deformed configurations for the swinging rod at different simulation times.}
\label{fig.pulsating.3}
\end{figure}

In Fig. \ref{fig.pulsating.3}, we present selected deformed configurations at several time instants over the $660$-second simulation. The figure consists of four subplots, corresponding to the piecewise definition of the pulsating driving force frequency. During the first time interval, $t \in [0, T_1]$, where the driving force frequency increases, we observe a sudden amplification of elastic vibrations along the rod. Simultaneously, the rod oscillates around a shifted static equilibrium position, thus exhibiting a combination of elastic vibrations and rigid body rotations, similar to those reported in \cite{Nguyen2024}. During the time interval $[T_1, T_2]$, where the driving force frequency remains constant, the rod exhibits reduced elastic vibrations while continuing to oscillate as a rigid body around a shifted static equilibrium position. However, its nonlinear dynamic behavior intensifies during $[T_2, T_3]$, where the driving force frequency decreases from $2$ Hz to zero. In this scenario, we observe the formation of a "bulge" located approximately at $\frac{1}{4}L$, along with a bending of the rod near its free end. This bulge is observed to move along the rod over time, thus indicating a sort of traveling wave phenomenon on the rod. This behavior is highly nonlinear and may arise from the complex interplay between the elastic properties of the rod, the surrounding flow, and the characteristics of the external driving force. 
However, a comprehensive understanding of these phenomena requires further investigation and lies beyond the scope of this study. Finally, when the external driving force \( \mathbf{f}(t) \) is removed (zero force frequency), all oscillations damp out (due to the surrounding flow), and the rod returns to its initial configuration.

\begin{figure}[h!]
\centering
\includegraphics[width=0.99\columnwidth]{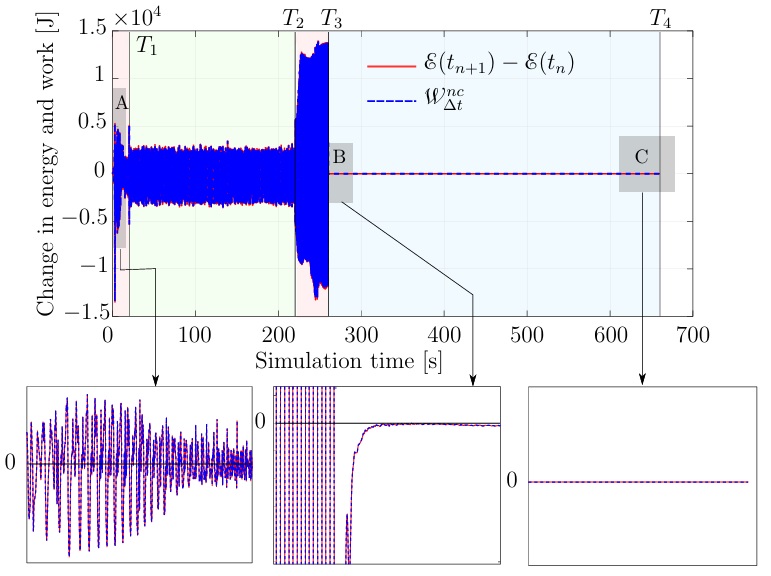}
\caption{Change in energy $\Delta\mathcal{E}$ of the swinging rod and work done by external and dissipative forces $\mathcal{W}_{\Delta t}^{nc}$.}
\label{fig.pulsating.4}
\end{figure}

In Fig. \ref{fig.pulsating.4}, we plot the change in total mechanical energy from $t = 0$ to $t = T_4$. The main objective of this graph is to show that the change in the system's total energy over a time step, $\Delta t = t_{n+1} - t_n$, matches the work performed by the external forces (pulsating + fluid forces), i.e. $\Delta\mathcal{E} = \mathcal{E}(t_{n+1})-\mathcal{E}(t_n)=\mathcal{W}_{\Delta t}^{nc}$, throughout the simulation. 

The expression for evaluating the work over a time step is derived from the definition of work done over a differential displacement, $d\curve(s,t)$ given by $d\mathcal{W}^{nc} = \mathbf{F}_{nc}(s,t) \cdot d\curve(s,t)$. After some algebraic manipulations, this expression can be rewritten as,
\begin{equation}\label{eq.pulsating.3}
    d\mathcal{W}^{nc}(t) = \int_0^L \left[ \mathbf{F}_{nc}(s,t) \cdot d\curved(s,t) \, ds \right] \, dt,
\end{equation}

\noindent
where $\mathbf{F}_{nc}$ collects all non-conservative forces acting along the rod. By integrating \eqref{eq.pulsating.3} from $t_n$ to $t_{n+1}$, we obtain the total work done by such non-conservative forces as follows,
\begin{equation}\label{eq.pulsating.4}
    \mathcal{W}_{t_n \longrightarrow t_{n+1}}^{nc} = \mathcal{W}_{\Delta t}^{nc} = \int_{t_n}^{t_{n+1}} \left[ \int_0^L \mathbf{F}_{nc}(s,t) \cdot d\curved(s,t) \, ds \right] \, dt.
\end{equation}

The time integration in \eqref{eq.pulsating.4} is approximated using the trapezoidal rule, leading to the following final expression for the work performed during a time step,
\begin{equation}\label{eq.pulsating.5}
\begin{split}
    \mathcal{W}_{\Delta t}^{nc} &\approx \frac{1}{2} \left\{ \left[ \int_0^L \mathbf{F}_{nc}(s) \cdot d\curved(s) \, ds \right]_{n+1} + \right. \\
    & \left. + \left[ \int_0^L \mathbf{F}_{nc}(s) \cdot d\curved(s) \, ds \right]_n \right\}.
    \end{split}
\end{equation}

In Fig. \ref{fig.pulsating.4}, we observe that the change in the total system energy closely matches the work done by the external forces, indicating that no artificial or numerical energy is introduced into the system. This balance can be seen through the whole simulation. In particular, at point B, when the pulsating driving force becomes zero, there is a significant drop in both the energy and the work. As expected, neither energy nor work reaches zero at this point due to the ongoing motion of the rod. Over time, as observed at point C, all oscillations are damped out, and the rod returns to its initial straight configuration. This behavior suggests that ARMoor is reliable and capable of accurately simulating mooring lines subjected to complex force systems, including following forces and time-varying driving forces. 

The small discrepancies between $\Delta\mathcal{E}$ and $\mathcal{W}_{\Delta t}^{nc}$ observed in Fig. \ref{fig.pulsating.4}, particularly when zoomed in, arise from the approximation of the integral in \eqref{eq.pulsating.5} and the selected simulation time step.

For ROD-D, we consider a mooring line immersed in still water, similar to the configuration used for ROD-B (see Subsec. \ref{subsec:cat}). The parameters associated with the surrounding flow and the barrier term (for simulating contact) are also identical to those used for ROD-B. For this example, we assumed a spatial discretization of $N_e=128$ elements and  time step $\Delta t = 0.01$ s. The rest solver settings are similar to the previous showcases. 

As outlined at the beginning of this subsection, this showcase involves two scenarios in which a pulsating driving force is applied at the fairlead, referred to as ROD-DT and ROD-DN. Prior to the application of the pulsating excitation, the fairlead is brought to its final position using a force control strategy over a duration of $T_1 = 30 \, \mathrm{s}$. The system is then allowed to relax for $T_2 = 80 \, \mathrm{s}$ to dissipate any undesired vibrations. Finally, the pulsating force is applied over a time windows of $T_3 = 80 \, \mathrm{s}$, thus resulting in a total simulation time of $T = T_1 + T_2 + T_3 = 190 \, \mathrm{s}$. To bring the fairlead to its final position prior applying the driving force, we use the data corresponding to case 3 from Table \ref{Cases.T3}.

The expressions for the pulsating force along the tangential direction, $\mathbf{f}_t$, and the normal direction, $\mathbf{f}_n$, to the rod centerline at the fairlead are given as follows,
\begin{equation}\label{eq.pulsating.6}
\begin{split}
    \mathbf{f}_t(t) &= A_F\sin(2\pi f t)\mathcal{P}_{\mathbf{d}_p}^{\parallel}(\mathbf{d}_p)=A_F\sin(2\pi f t)\mathbf{d}_p,\quad\text{and} \\
    \mathbf{f}_n(t) &= A_F\sin(2\pi f t)\mathcal{P}_{\mathbf{d}_p}^{\perp}(\mathbf{d}_p),
    \end{split}
\end{equation}

\noindent
where the force amplitude is assumed to be constant with a value of $A_F = 525$ kN. The rod director $\mathbf{d}_p$ represents the orientation of the rod at the moment when the pulsating force is applied, i.e. $\mathbf{d}_p(T_2)$, and remains fixed throughout the rest of the simulation. The term $\mathcal{P}_{\mathbf{d}_p}^{\perp}(\mathbf{d}_p)$ denotes the normal direction to the centerline at the same temporal point (see Fig. \ref{fig.pulsating.1}). It is important to note that computing the normal direction in this manner is feasible due to the two-dimensional nature of the problem. In a three-dimensional context, an infinite number of normal directions would be possible. Since both $\mathbf{d}_p$ and $\mathcal{P}_{\mathbf{d}_p}^{\perp}(\mathbf{d}_p)$ remain constant during the interval $[T_2,T_3]$, the resulting pulsating forces are not follower forces. To investigate the dynamic behavior of the mooring line under the conditions mentioned above, we consider frequencies ranging from $0.25$ Hz to $6.0$ Hz.

\begin{figure}[h!]
\centering
\includegraphics[width=0.99\columnwidth]{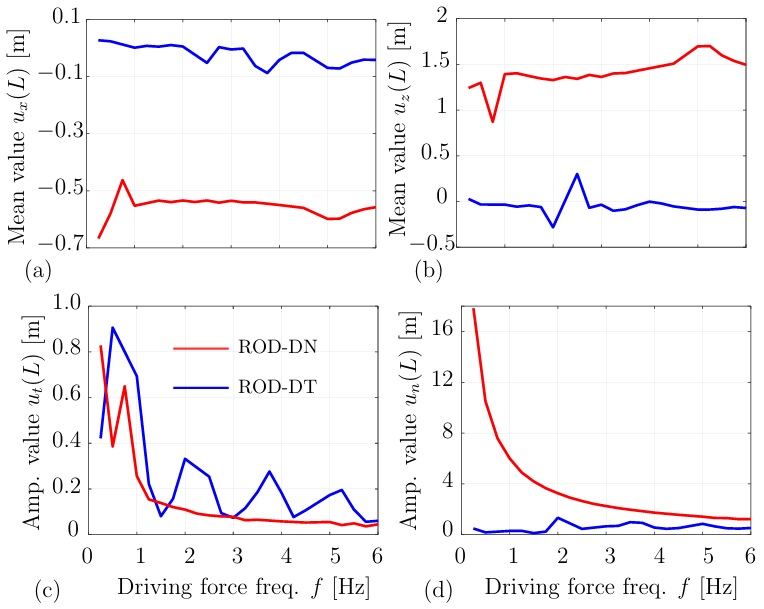}
\caption{(a)-(b) Mean value of the fairlead displacement $u_x(L)$ and $u_z(L)$. (c)-(d) Amplitude value of the fairlead displacement along the tangential direction $\mathbf{d}_p$ and normal direction $\mathcal{P}_{\mathbf{d}_p}(\mathbf{d}_p)$.}
\label{fig.pulsating.5}
\end{figure}

In Fig. \ref{fig.pulsating.5}a-b, we present the mean displacement of the fairlead along the $X$- and $Z$-directions, $u_x(L)$ and $u_z(L)$, respectively, as a function of the driving force frequency, $f$, after reaching steady state. Conversely, Fig. \ref{fig.pulsating.5}c-d illustrates the amplitude of the fairlead displacement along the tangential and normal directions relative to $\mathbf{d}_p$, denoted as $u_t(L)$ and $u_n(L)$, respectively, once steady state is attained. These quantities were computed by analyzing the last 10 seconds of the simulation. On one hand, displacements $u_x$ and $u_z$ are obtained as follows,
\begin{equation}\label{eq.pulsating.7}
\begin{split}
    & u_k(L,t) = \varphi_k(L,t) - \overline{\varphi}_k, \\
    & \text{with}\quad\overline{\varphi}_k = \frac{1}{n}\sum_{j=1}^n \varphi_k(L,t_j),
    \end{split}
\end{equation}

\noindent
where $k$ represents either $x$ or $z$, and $n$ denotes the number of points used to compute the averaged position of the fairlead during the final portion of the dynamical relaxation time window, $T_2$, prior applying the pulsating load. 
On the other hand, displacements along the tangential and normal direction at the fairlead, $u_t$ and $u_n$, are obtained by performing a coordinate transformation of $\mathbf{u}=(u_x,0,u_z)^T$ from the global frame $\mathcal{E}$ to a local frame at the fairlead aligned with $\mathbf{d}_p$ and $\mathcal{P}_{\mathbf{d}_p}^{\perp}(\mathbf{d}_p)$. 

For the case ROD-DN, the mean values of $\varphi_x(L)$ and $\varphi_z(L)$ exhibit slight oscillations at low driving force frequencies (between $0.25$ Hz and $1$ Hz), remaining nearly constant up to $5$ Hz, where additional oscillatory behavior becomes evident. As it can be seen, the mean displacement of the fairlead shifts toward the anchor by approximately $0.5$ m in the $X$-direction and upward by approximately $1.5$ m in the $Z$-direction. In contrast, for ROD-DT, the mean displacement along both directions oscillates closely around zero, indicating no significant shift in the mean value but rather a slight fluctuation around it.

\begin{figure}[h!]
\centering
\includegraphics[width=0.99\columnwidth]{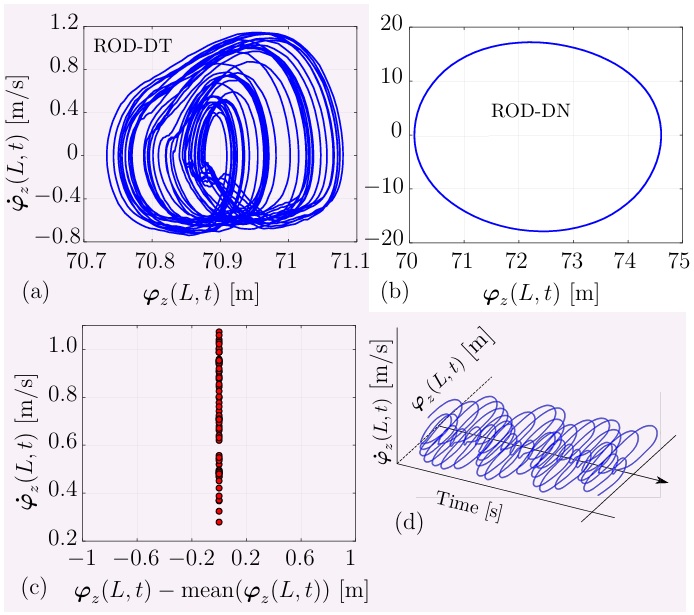}
\caption{(a)-(b) Fairlead phase portrait for $f=1.25$ Hz. (c) Poincaré map plot related to the phase portrait (a). (d) Three-dimensional phase-space plot for ROD-DT.}
\label{fig.pulsating.6}
\end{figure}

Regarding the amplitude of the fairlead displacement, case ROD-DN exhibits a monotonic decrease as the driving force frequency increases. For $\varphi_t(L)$, some oscillatory behavior is observed at low frequencies, whereas $\varphi_n(L)$ shows a consistent monotonic decrease across the entire range of investigated frequencies. Furthermore, it is evident that the amplitude of both $\varphi_t(L)$ and $\varphi_n(L)$ scales inversely with $f^2$ at higher frequencies. This behavior suggests that, when the mooring line is subjected to a pulsating force applied in the normal direction to the centerline at the fairlead, it responds in a manner analogous to a classical damped mass-spring system. To elucidate this point, let us consider a damped harmonic oscillator with mass $m$, spring stiffness $k$, and damping coefficient $b$, under the action of a sinusoidal driving force of the form $A_0\sin(\Omega_0 t)$. According to classical vibration theory, the steady-state response of this linear system is given by,
\begin{equation}\label{eq.pulsating.8}
\begin{split}
    A = \frac{A_0}{\sqrt{m(\Omega_0^2-\omega_0^2)^2 + b^2\Omega_0^2}}, 
\end{split}
\end{equation}

\noindent
where $\omega_0 = \sqrt{k/m}$ denotes the natural frequency of the system. For $\Omega_0 > \omega_0$, it is well established that the amplitude $A$ in \eqref{eq.pulsating.8} decreases as $\Omega_0$ increases. Within the framework of this work, the responses of ROD-DN are consistent with the predictions of linear vibration theory. Additionally, Fig. \ref{fig.pulsating.6}b illustrates the phase portrait at the fairlead point for $f=1.25$ Hz along the $Z$-coordinate, revealing a classical closed trajectory indicative of periodic motion. Similar phase portraits are obtained for the $X$-coordinate and for different force frequencies.

For case ROD-DT, the amplitudes of $u_t$ and $u_n$ exhibit an oscillatory behavior as the pulsating force frequency increases. Moreover, for ROD-DN, there is an order-of-magnitude difference between the amplitudes of $u_t$ and $u_n$, whereas for ROD-DT, these amplitudes are of comparable magnitudes. This finding suggests that excitation in the normal direction minimally affects the axial dynamics, indicating that it is predominantly dominated by first-order effects. In contrast, for ROD-DT, we observe a strong coupling between axial and bending dynamics, implying that tangential loads lead to second-order dynamic effects. 

This assertion is further supported by the significant differences observed in the phase portraits of the fairlead point when subjected to tangential versus normal pulsating forces at $f = 1.25$ Hz. As noted earlier, the phase portrait for ROD-DN is characterized by an elliptical shape, indicative of periodic motion, while ROD-DT exhibits a far more intricate trajectory (see Fig. \ref{fig.pulsating.6}a). Such a phase portrait shows curves that intersect themselves multiple times, a phenomenon associated with \textit{period-$k$ motion}. However, due to the apparent complexity of the trajectory and the overlapping curves, it is not immediately evident that the motion remains periodic. To further investigate this, we compute the Poincaré map of the phase portrait by choosing as the Poincaré section $\varphi_z(L,t) - \overline{\varphi}_z(L,t) = 0$ (see Fig. \ref{fig.pulsating.6}c). The resulting map reveals that the trajectories crossing this section organize into two distinct clusters, suggesting a period-2 motion. This allows us to infer that the motion of the mooring line is still periodic, with the trajectory repeating only after two cycles of the driving force. It is worth noting, however, that the system may be undergoing a transition to a higher period-$k$ motion state. This is hinted by the slightly scattered points in the lower cluster of the Poincaré map and the intricate phase portrait of ROD-DT. Such behavior is characteristic of systems approaching chaotic dynamics. For example, as system parameters vary (e.g. driving force frequency), the motion may undergo period-doubling bifurcations (from period-2 to period-4, period-8, etc.), eventually leading to chaos. 
To provide additional insight into the complexity of the motion, we build a three-dimensional phase-space representation of the phase portrait by incorporating time as an additional axis (see Fig. \ref{fig.pulsating.6}d). This visualization highlights the intricate nature of the fairlead trajectory. 

Nonetheless, a comprehensive characterization of the nonlinear dynamics of mooring lines under tangential loads at the fairlead, particularly the potential onset of chaos, lies beyond the scope of the current work.

\begin{figure}[h!]
\centering
\includegraphics[width=0.99\columnwidth]{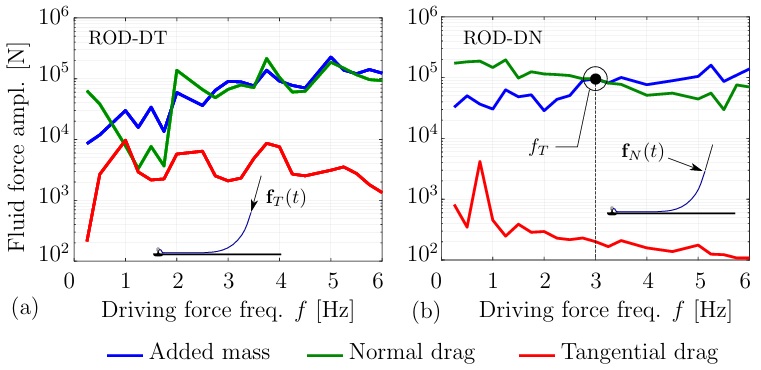}
\caption{Amplitude of three force components induced by the surrounding flow at rest, integrated over the mooring line at steady state.}
\label{fig.pulsating.7}
\end{figure}

In Fig. \ref{fig.pulsating.7}, we illustrate the amplitude $A_f^{\Omega}$ of three components of the resulting force induced by the surrounding flow (added mass, normal drag, and tangential drag), integrated over the mooring line length once the rod reaches the steady state, that is,
\begin{equation}\label{eq.pulsating.9}
\begin{split}
    A_f^{\Omega} &= \underset{\text{steady st.}}{\max}\left(F_f^k(t)\right) - \underset{\text{steady st.}}{\min}\left(F_f^k(t)\right),\quad\text{and} \\
    F_f^k(t) &= \int_0^L \abss{\mathbf{F}_f^k(s,t)}\,ds,
\end{split}
\end{equation}

\noindent
where $k$ stands for the added mass, tangential drag or the normal drag component (see \cite{Nguyen2024}). We observe that the tangential drag component presents an oscillatory behavior as a function of the pulsating driving force for case ROD-DT, when compared with case ROD-DN, where it shows a decreasing behavior as $\Omega_0$ increases. It can be explained based on the fact that tangential loads directly affect the axial dynamics of the rod thus producing more tangential drag dissipation when compared to ROD-DN like scenarios. However, for both ROD-DN and ROD-DT, the tangential drag is an order of magnitude smaller than the added mass and normal drag components. 

Moreover, for case ROD-DN, we observe different trends in the contributions of the added mas and normal drag as the driving force frequency increases. For low driving frequencies, the normal drag dominates the fluid force. This is attributed to the relatively slow motion of the mooring line, which leads to moderate relative velocities with respect to the surrounding fluid. In this regime, the drag force, which scales with the square of the relative velocity $\abss{\mathbf{V}}^2$, outweighs the added mass term. As the frequency of the pulsating force increases, the added mass term becomes the dominant component of the fluid force. This transition is due to the rapid accelerations of the mooring line at higher frequencies, which enhance the inertial resistance imposed by the surrounding fluid. Since the added mass force scales with the acceleration, which increases with the square of the frequency $f$, it overrides the normal drag at higher driving frequencies. This transition from a drag-dominated regime at low frequencies to an added-mass-dominated regime at higher frequencies is illustrated in Fig. \ref{fig.pulsating.7}b (black dot at $f=3$ Hz).

Conversely, for ROD-DT, both the added mass and normal drag forces increase as the driving force frequency rises. Although the increase in added mass is a direct consequence of larger accelerations at high frequencies, the normal drag increases due to the induced motion in directions normal to the centerline, which in turn increases the relative velocity between the fluid and the mooring line.

These findings highlight the frequency-dependent interplay between added mass and drag forces in the dynamics of flexible mooring lines. While the dominance of the added mass term at high frequencies is consistent with classical fluid-structure interaction theory, the behavior under tangential forcing underscores the nonlinear coupling between axial and bending dynamics.

\begin{figure}[h!]
\centering
\includegraphics[width=0.99\columnwidth]{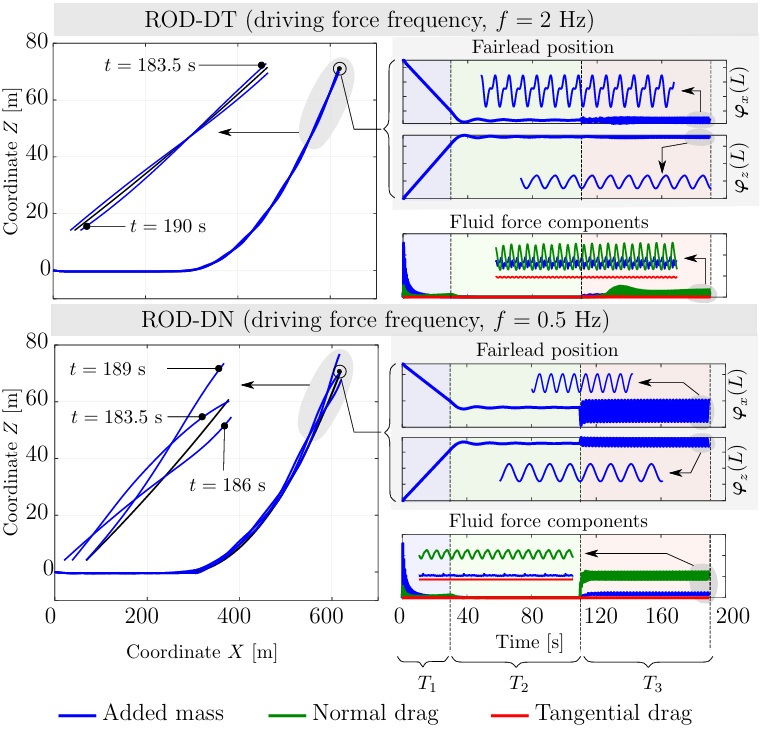}
\caption{Deformed configurations for cases ROD-DT and ROD-DN at $f=2.0$ Hz and $f=0.5$ Hz, respectively.}
\label{fig.pulsating.8}
\end{figure}

Finally, in Fig. \ref{fig.pulsating.8}, we present selected deformed configurations at different time instants for case ROD-DT at $f=2$ Hz and case ROD-DN at $f=0.5$ Hz. As observed, the deformed configurations corresponding to ROD-DT exhibit smaller transverse oscillations, consistent with conclusions drawn from Fig. \ref{fig.pulsating.5}, while ROD-DN shows larger oscillations in the transverse direction around a shifted static equilibrium configuration. Additionally, we also depict the position time series along the $X$ and $Z$ directions for both ROD-DT and ROD-DN.

Regarding the fluid force components, it is evident that, for both cases, the normal drag dominates over the added mass term. This observation aligns with the analysis presented earlier, as the driving force frequency for ROD-DN is below the transition point from a drag-dominated regime to an added-mass-dominated regime.



\subsection{Case 3: UMaine VolturnUS-S marine platform}\label{subsec:case3}

\begin{figure}[h!]
\centering
\includegraphics[width=0.99\columnwidth]{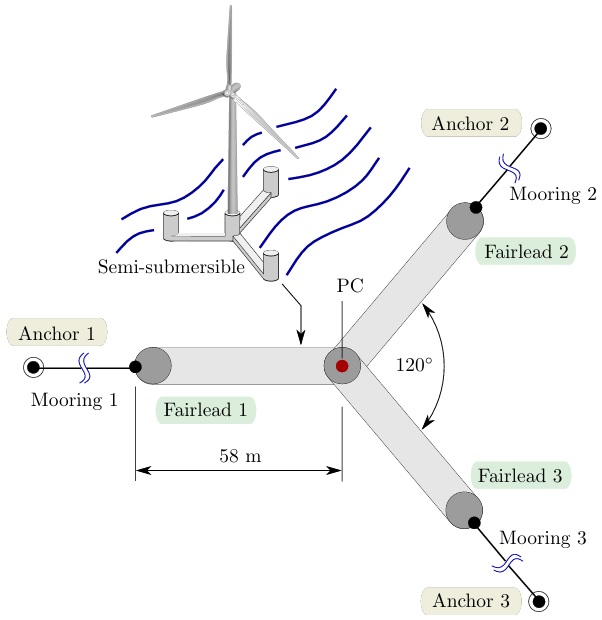}
\caption{Mooring setup for the VolturnUS-S.}
\label{fig.mooringsetup}
\end{figure}

In this subsection, we present a comprehensive comparison between our ARMoor model and OpenFAST for the mooring system response of a floating offshore wind turbine. To this end, we used the IEA 15MW wind turbine \cite{gaertner2020} on the VolturnUS-S marine platform \cite{allen2020} under step-wind loading conditions. The sub-structure consists of a three-legged semi-submersible platform with a central pillar where the wind turbine is placed (see Fig. \ref{fig.mooringsetup}).   

The VolturnUS-S mooring system configuration consists of a three-line configuration spaced at an angular separation of $120^{\circ}$ from each other. In Table \ref{Mooringsetup.T6}, we list the coordinates of fairleads and anchors with respect to the platform center (point PC in Fig. \ref{fig.mooringsetup}). The VolturnUS-S mooring system is based on a chain-type catenary setup ($185$ mm diameter chain links) connected to the substructure at a depth of $14.0$ m below the sea level. The mooring line has an axial stiffness of $EA=3270$ MN and dry density per unit length of $A_{\rho}=685$ kg/m. For dynamic simulations, the bending stiffness is set to zero ($EI = 0$); however, this approach is not applicable for static simulations, as the strain operator becomes singular in the absence of bending stiffness. The fairlead pretension is $2437$ kN at a fairlead angle of $56.4^{\circ}$. As a result, the total vertical force due to the mooring system is $6084$ kN (or $620$ tons). 

A full description of the dataset associated with the VolturnUS-S platform is available in \cite{allen2020}. Although VolturnUS-S is chain-based, it constitutes an excellent mooring system for comparison purposes. 

As a previous step to the comparison analysis between models, we ran six verification tests with OpenFAST v.3.5.0 in order to replicate the results obtained by the University of Maine regarding the VolturnUS-S platform \cite{allen2020}. Such tests are properly reported in \cite{Veseth2023} and consist of an analysis of natural frequencies and damping ratios, in a wind-free and wave-free environment, for all 6-degree-of-freedom of the floating platform.    

\begin{table}[h]
\centering
\caption{Fairlead and anchor coordinates.}
\label{Mooringsetup.T6}
\begin{tabular}{ >{\arraybackslash}m{.18\linewidth} 
>{\centering\arraybackslash}m{.14\linewidth} 
>{\centering\arraybackslash}m{.18\linewidth} >{\centering\arraybackslash}m{.22\linewidth}}
\hline\hline
\multirow{2}{*}{Mooring line} & \multirow{2}{*}{Length [m]} & \multicolumn{2}{c}{Coordinates [m]} \\
\cline{3-4}
\multicolumn{1}{l}{} & & Fairlead & Anchor \\
\hline
\multirow{3}{*}{Line 1} & \multirow{3}{*}{850.0} & $x=-58.0$ & $x=-837.60$ \\
\multicolumn{1}{l}{} & & $y=0.00$ & $y=0.00$ \\
\multicolumn{1}{l}{} & & $z=-14.0$ & $z=-200.0$ \\
\hline
\multirow{3}{*}{Line 2} & \multirow{3}{*}{850.0} & $x=29.0$ & $x=418.8$ \\
\multicolumn{1}{l}{} & & $y=50.2$ & $y=725.4$ \\
\multicolumn{1}{l}{} & & $z=-14.0$ & $z=-200.0$ \\
\hline
\multirow{3}{*}{Line 3} & \multirow{3}{*}{850.0} & $x=29.0$ & $x=418.8$ \\
\multicolumn{1}{l}{} & & $y=-50.2$ & $y=-725.4$ \\
\multicolumn{1}{l}{} & & $z=-14.0$ & $z=-200.0$ \\
\hline\hline
\end{tabular}
\end{table}

As simulation engine for the mooring system, we use MoorDyn, an open-source, lumped-mass mooring dynamics model integrated into OpenFAST for simulating the behavior of mooring lines, which accounts for factors like hydrodynamic forces and seabed interaction. To compare the response provided by MoorDyn and that obtained using our current model, we proceed as follows:
\begin{enumerate}
    \item Run OpenFAST + MoorDyn for the floating wind turbine (IEA15 MW + VolturnUS-S platform).
    \item Extract the force time-series at the fairlead positions from OpenFAST simulations.
    \item Interpolate the force time-series to match the time step required by our rod model. 
\end{enumerate}

Here, we conduct two analyses: \textit{i}) a dynamic relaxation study employing both OpenFAST and our rod model (referred to hereafter as ROD-E) to compare their resulting static equilibrium positions in a wind-free and wave-free environment, and \textit{ii}) a comparative analysis between OpenFAST and our model under a step-wind load scenario (hereafter referred to as ROD-F).

For ROD-E, we let run OpenFAST + MoorDyn over $7200$ s. To filter out potential start-up transients and allow the motion of the floating platform to dampen as much as possible, we discard the first $6600$ s of simulation. The final configuration of the mooring line and the fairlead tension are then determined by averaging their values over the last $600$ s. Table \ref{Vultur.T1} lists the average position and tension for all three fairleads as obtained from OpenFAST.   

\begin{table}[h!]
\centering
\caption{ROD-E - OpenFAST dynamic relaxation simulation.}
\label{Vultur.T1}
\begin{tabular}{ >{\arraybackslash}m{.19\linewidth} 
>{\centering\arraybackslash}m{.16\linewidth} 
>{\centering\arraybackslash}m{.16\linewidth} >{\centering\arraybackslash}m{.16\linewidth}}
\hline\hline
Solution & Fairlead 1  & Fairlead 2  & Fairlead 3 \\ 
\hline
$x(L)$ [m] & -57.23 & 29.75 & 29.75  \\ [0.1cm]
$y(L)$ [m] & 0.00 & 50.23 & -50.23 \\[0.1cm]
$z(L)$ [m] & -15.83 & -13.61 & -13.61 \\[0.1cm]
$N_x(L)$ [kN] & 1340.65 & -670.33 & -670.33 \\[0.1cm]
$N_y(L)$ [kN] & 0.00 & -1163.27 & 1163.27 \\[0.1cm]
$N_z(L)$ [kN] & 2010.35 & 2026.96 & 2026.96 \\[0.1cm]
\hline\hline
\end{tabular}
\end{table}

While a static analysis is the appropriate approach to determine the final static configuration and fairlead tension, we conducted a dynamic relaxation study in OpenFAST to assess the accuracy of our current model in comparison with OpenFAST under similar conditions. It is worth noting that the results obtained using dynamic relaxation in OpenFAST closely approximate those from a static analysis performed within the same software \cite{Veseth2023}. 

For this case, ARMoor is run for 1800 s using a force control strategy. The force applied at the fairlead corresponds to the average tension from OpenFAST, as reported in Table \ref{Vultur.T1}. Our focus is on mooring line 1, which is discretized into $N_e = 200$ elements. For this showcase, we assumed the following settings for the solver: time step $\Delta t=0.1$, polynomial degree $p=3$, continuity $r=1$, and tolerance error for Newton iterations $\varepsilon_{tol}=10^{-10}$.

\begin{figure}[h!]
\centering
\includegraphics[width=0.99\columnwidth]{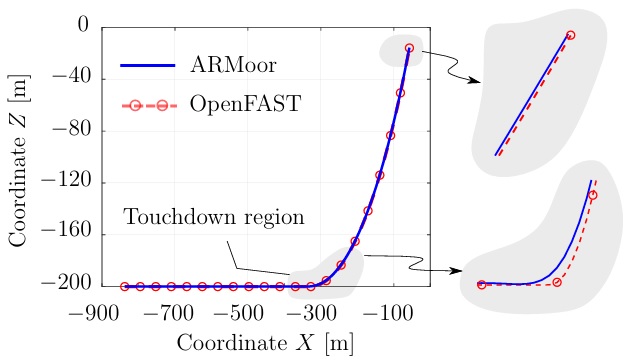}
\caption{Comparison between OpenFAST and ARMoor - dynamic relaxation simulation.}
\label{fig.ARMoor_OpenFAST}
\end{figure}

In Fig. \ref{fig.ARMoor_OpenFAST}, we present the final ``static'' configuration and tension distribution along mooring line 1, as obtained from both ARMoor and OpenFAST. The final configuration of the line from our rod model closely aligns with the OpenFAST results. We also observe that our solution is stiffer compared to the results obtained from OpenFAST. This behavior is consistent with the fact that OpenFAST employs a lumped spring-damper-mass model, which is better suited for chain-based analyses. Although the bending stiffness ($EI$) of ARMoor was set to zero in this case, the model inherently incorporates geometric constraints that enforce curvature continuity. As a result, it facilitates smoother curvature transitions, leading to a ``stiffer'' response of the mooring line compared to lumped-parameter models.


Table \ref{Vultur.T2} presents the final position of fairlead 1 as obtained with ARMoor after allowing the system to dynamically relax to its static equilibrium. The values reported correspond to the original discretization of the mooring line ($N_e=200$), with a polynomial degree of $p=5$ and continuity order $r=2$. Comparison with OpenFAST results reveals a maximum difference of $0.73\%$, indicating excellent agreement with this well-established code in the wind energy sector.

\begin{table}[h!]
\centering
\caption{Fairlead 1 position - comparison between ARMoor and OpenFAST.}
\label{Vultur.T2}
\begin{tabular}{ >{\arraybackslash}m{.22\linewidth} 
>{\centering\arraybackslash}m{.15\linewidth} 
>{\centering\arraybackslash}m{.15\linewidth} >{\centering\arraybackslash}m{.19\linewidth}}
\hline\hline
Position [m] & OpenFAST  & ARMoor  & $\overline{\text{Diff.}}_X(\cdot,\cdot)$ $[\%]$ \\ 
\hline
$x$-coordinate & -57.23 & -57.510 & 0.489  \\ [0.1cm]
$y$-coordinate & 0.00 & 0.00 & 0.00 \\[0.1cm]
$z$-coordinate & -15.83 & -15.713 & 0.739 \\[0.1cm]
\hline\hline
\end{tabular}
\end{table}



\begin{figure}[h!]
\centering
\includegraphics[width=0.99\columnwidth]{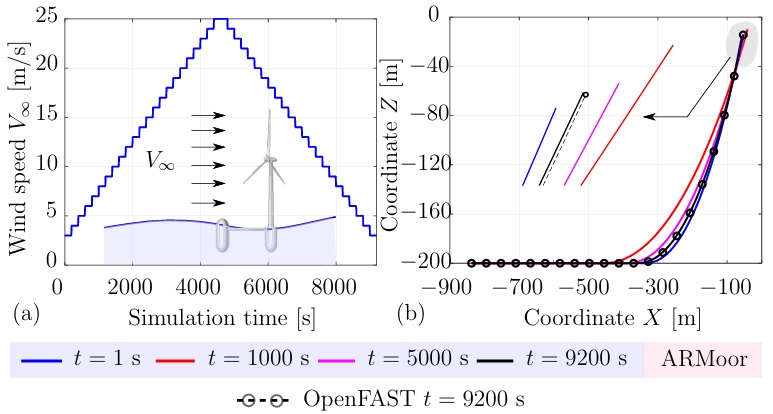}
\caption{(a) Free-stream speed over time. (b) Snapshots of the mooring line 1 at different time steps obtained using ARMoor.}
\label{fig.ARMoor_1}
\end{figure}

For ROD-F, we consider a wave-free environment with a uniform, stepped wind profile. The wind speed increases in $200$ s intervals from $3$ m/s to $25$ m/s and then decreases back to $3$ m/s (see Fig. \ref{fig.ARMoor_1}a). Solver settings are consistent with those used for ROD-E. As described previously, we begin by running OpenFAST for $9200$ s, after which we extract time-series forces at each fairlead to use as input in ARMoor via a force-control approach. Given that OpenFAST and ARMoor use different time steps, the forces from OpenFAST are thus interpolated to align with ARMoor's time step requirements. Additionally, ARMoor requires force data at the midpoint of each time step, i.e., at $n+\frac{1}{2}$.

In Fig. \ref{fig.ARMoor_1}b, we present a series of snapshots depicting the configuration of Mooring Line 1 at various time instants. Notably, the configuration of the ARMoor solution at $t = 9200$ s shows excellent agreement with that obtained using OpenFAST. The mean absolute difference along the $X$- and $Z$-coordinates between the two models at $t=9200$ s is $\text{MAD}_X=0.1691$ and $\text{MAD}_Z=0.1194$, respectively, indicating that the ARMoor and OpenFAST solutions align closely, with only minor deviations. In other words, the predicted values by ARMoor deviate, on average, by $17$ cm along $X$ and $12$ cm along $Z$ from the results obtained via OpenFAST. Similar results were obtained for the other two mooring lines. For comparison, the MAD was computed using nodal values from both models.

\begin{table}[h!]
\centering
\caption{ROD-F - Tension at the clamped end.}
\label{Vultur.T3}
\begin{tabular}{ >{\arraybackslash}m{.18\linewidth} 
>{\centering\arraybackslash}m{.23\linewidth} 
>{\centering\arraybackslash}m{.2\linewidth} >{\centering\arraybackslash}m{.19\linewidth}}
\hline\hline
Mooring line & OpenFAST [MN]  & ARMoor [MN]  & $\overline{\text{Diff.}}_N(\cdot,\cdot)$ $[\%]$ \\ 
\hline
Line 1 & 1.5811 & 1.5863 & 0.3258  \\ [0.1cm]
Line 2 & 1.2366 & 1.2490 & 1.0072 \\[0.1cm]
Line 3 & 1.2398 & 1.2521 & 0.9906 \\[0.1cm]
\hline\hline
\end{tabular}
\end{table}

Regarding the reaction forces at the clamped end, Table \ref{Vultur.T3} presents the tension at $s=0$ for the three mooring lines, as computed by both models. To mitigate the effects of oscillatory peaks and potential phase shifts—arising from the dynamic nature of the study and the differing model responses—we calculated the mean tension over the last 600 time steps. As shown in Table \ref{Vultur.T3}, the maximum tension deviation between both models is $1\%$, exhibiting excellent agreement not only in the final configuration shape but also in the prediction of reaction forces, despite the fundamentally different modeling approaches employed.

\begin{figure}[h!]
\centering
\includegraphics[width=0.99\columnwidth]{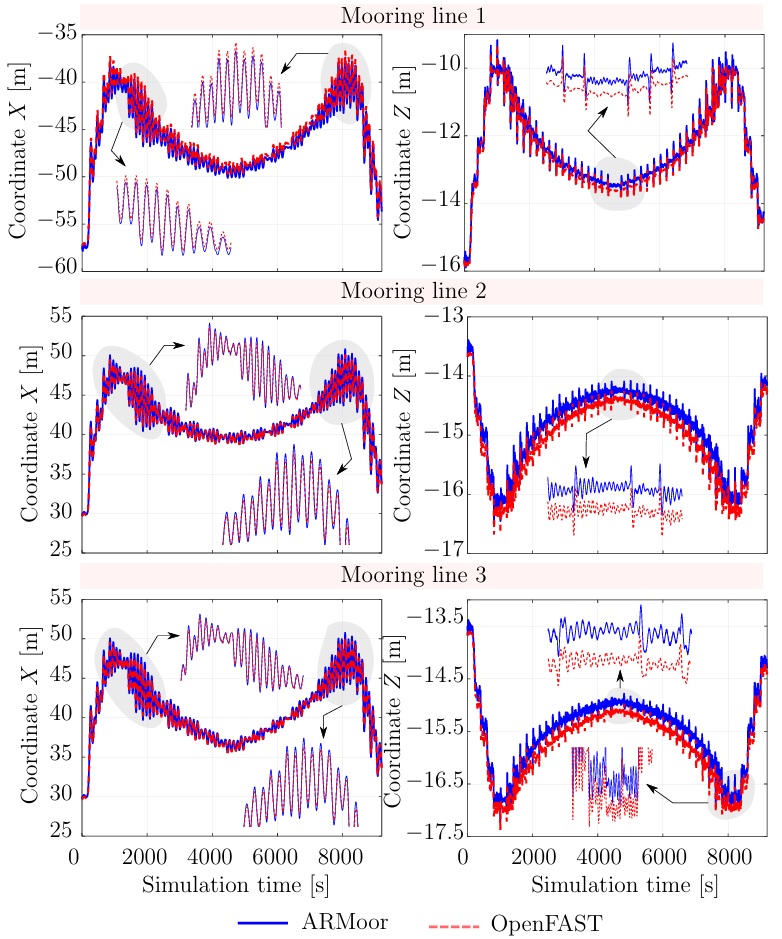}
\caption{Fairlead position (node 200) over time.}
\label{fig.ARMoor_OpenFAST_2}
\end{figure}

\begin{table}[h!]
\centering
\caption{ROD-F - mean absolute difference for the fairlead position.}
\label{Vultur.T4}
\begin{tabular}{ >{\arraybackslash}m{.18\linewidth} 
>{\centering\arraybackslash}m{.23\linewidth} 
>{\centering\arraybackslash}m{.24\linewidth}}
\hline\hline
Mooring line & $\text{MAD}_{200,t}(\cdot,\cdot)$ [m] & $\text{MAD}_{200,t}(\cdot,\cdot)$ [m] \\ 
\hline
Line 1 & 0.4753 & 0.1280 \\ [0.1cm]
Line 2 & 0.1850 & 0.1977 \\ [0.1cm]
Line 3 & 0.1853 & 0.2102 \\ [0.1cm]
\hline\hline
\end{tabular}
\end{table}

Figure \ref{fig.ARMoor_OpenFAST_2} presents the temporal evolution of the fairlead position ($X$ and $Z$ coordinates) for all three mooring lines. Across all plots, the agreement between the ARMoor and OpenFAST solutions is generally very good, with slightly larger deviations observed in the $Z$-coordinate for mooring lines 2 and 3. Conversely, for mooring line 1, the larger deviation is observed in the $X$-coordinate. To quantify these deviations, we computed the mean absolute difference at Node 200 (fairlead) over the entire simulation time $\text{MAD}_{200,t}(\cdot,\cdot)$ for all three mooring lines. For mooring line 1, the ARMoor solution shows an average deviation of $47$ cm from the OpenFAST solution. Despite this deviation, a maximum MAD of approximately 47 cm (see Table \ref{Vultur.T4}) over a characteristic length of $850$ m (mooring line length) is relatively small. This result reaffirms the high accuracy of our model when compared to well-established and widely recognized tools such as OpenFAST.

It should be noted that the temporal evolution of the fairlead position along the $Z$-coordinate for mooring lines 2 and 3 exhibits smaller amplitudes compared to those obtained with OpenFAST, despite the frequency and phase aligning closely. This behavior can be attributed to the fact that OpenFAST solves the coupled aero-hydro-elastic problem, accounting for the interactions between the mooring system and the entire wind turbine. In contrast, ARMoor focuses solely on the mooring line dynamics, using the force at the fairlead provided by OpenFAST as input. Consequently, the pitch and heave motions of the floater might introduce additional fairlead displacements in the vertical direction, thereby explaining the observed differences in the plots.


\section{Limitations of the current model}\label{sec:limitations}

Although the numerical results obtained with the present framework have been verified using several well-established solutions from the literature and industry-standard codes, there are limitations related to the current work that warrant discussion. 

First, ARMoor, in its current form, does not support arbitrary cross-sections or composite materials. This limitation arises naturally from the assumed configuration space $\mathcal{D}$ for the rod. Specifically, the rod's configuration is described solely by a curve $\curve(s) \in \mathbb{R}^3$ leading inherently to a torsion- and shear-free formulation. Second, the present study considers only linear elastic material behavior, which may significantly restrict the applicability and accuracy of the rod model when predicting the real-world performance of mooring lines. As part of future work, we aim to address this limitation by incorporating viscoplasticity into the formulation of the advanced nonlinear Kirchhoff rod model utilized in this study.

Finally, it is important to note that the computational cost of simulations with ARMoor is higher compared to lumped-parameter models, such as those implemented in OpenFAST or OrcaFlex.


\section{Concluding remarks}\label{ConcluSec}

In this paper, we explored the application of the nonlinear rod formulation presented in \cite{gebhardt2021} for rods undergoing only axial and bending deformations to analyze mooring lines in the field of offshore wind energy.

First, we extended the formulation in \cite{gebhardt2021} by incorporating interactions between the mooring line and the seabed. In addition, we enhanced the fluid model by including the buoyancy force and linear terms associated with the tangential and normal drag. The ``contact'' between the line and the seafloor is simulated by enhancing the system's Lagrangian with a penalization term based on barrier functions. Specifically, we examined the behavior of various barrier functions and their suitability for mooring system simulations. The implemented fluid model, while simplified, accounts for key forces, including added mass (fluid inertia effects), tangential and normal drag forces, and buoyancy.

Second, we verified the numerical framework against well-established solutions from the literature, such as the elastic catenary, as well as results obtained using OpenFAST, a widely recognized tool in the wind energy sector. Moreover, ARMoor's framework is employed to investigate the dynamic response of mooring lines subjected to pulsating driving forces applied at the fairlead.

In summary, this study addresses the application of a nonlinear rod formulation to investigate the static and nonlinear dynamic behavior of mooring line systems under a variety of loading conditions. To the best of the authors’ knowledge, the utilization of the formulation in \cite{gebhardt2021}, along with the results and discussions presented throughout this article, represents a novel contribution to the context of offshore wind energy applications.


\clearpage
\appendix

\section{Solution tables for ROD-B}\label{appeA}

\setcounter{table}{0}

\begin{table}[h!]
\centering
\caption{ROD-B (case 3) - Displacement control.}
\label{Case3.TA1}
\begin{tabular}{ >{\arraybackslash}m{.35\linewidth} 
>{\centering\arraybackslash}m{.14\linewidth} 
>{\centering\arraybackslash}m{.18\linewidth} >{\centering\arraybackslash}m{.14\linewidth}}
\hline\hline
Solution & Ref. sol. \cite{Gunnar2024}  & ARMoor ($N_e=128$)  & $\overline{\text{Diff.}}_X(\cdot,\cdot)$ $[\%]$ \\ 
\hline
$x_{touch}$ [m] & 277.5549 & 267.0753 & 3.775  \\ [0.1cm]
$\theta(L)$ [$^{\circ}$] & 22.9632 & 22.9199 & 0.188 \\[0.1cm]
Stretched length, $L_f$ [m] & 628.45 & 628.4419 & 0.046 \\[0.1cm]
$N_x(L)$ [kN] & 2030.303 & 2024.443 & 0.288 \\[0.1cm]
$N_z(L)$ [kN] & 860.2739 & 857.764 & 0.291\\[0.1cm]
\hline\hline
\end{tabular}
\end{table}

\begin{table}[H]
\centering
\caption{ROD-B (case 6) - Displacement control.}
\label{Case6.TA2}
\begin{tabular}{ >{\arraybackslash}m{.35\linewidth} 
>{\centering\arraybackslash}m{.14\linewidth} 
>{\centering\arraybackslash}m{.18\linewidth} >{\centering\arraybackslash}m{.14\linewidth}}
\hline\hline
Solution & Ref. sol. \cite{Gunnar2024}  & ARMoor ($N_e=128$)  & $\overline{\text{Diff.}}_X(\cdot,\cdot)$ $[\%]$ \\ 
\hline
$x_{touch}$ [m] & 86.17 & 85.1118 & 1.228  \\ [0.1cm]
$\theta(L)$ [$^{\circ}$] & 14.9079 & 14.8900 & 0.120 \\[0.1cm]
Stretched length, $L_f$ [m] & 630.55 & 630.5458 & 6.66$\times 10^{-4}$ \\[0.1cm]
$N_x(L)$ [kN] & 5000 & 4970.897 & 0.582 \\[0.1cm]
$N_z(L)$ [kN] & 1331.1354 & 1321.483 & 0.725 \\[0.1cm]
\hline\hline
\end{tabular}
\end{table}

\begin{table}[H]
\centering
\caption{ROD-B (case 3) - Force control.}
\label{Case3.TA3}
\begin{tabular}{ >{\arraybackslash}m{.35\linewidth} 
>{\centering\arraybackslash}m{.14\linewidth} 
>{\centering\arraybackslash}m{.18\linewidth} >{\centering\arraybackslash}m{.14\linewidth}}
\hline\hline
Solution & Ref. sol. \cite{Gunnar2024}  & ARMoor ($N_e=128$)  & $\overline{\text{Diff.}}_X(\cdot,\cdot)$ $[\%]$\\ 
\hline
$x_{touch}$ [m] & 277.5549 & 267.0815 & 3.773  \\ [0.1cm]
$\theta(L)$ [$^{\circ}$] & 22.9632 & 22.9293 & 0.147 \\[0.1cm]
Stretched length, $L_f$ [m] & 628.45 & 628.4561 & 9.70$\times 10^{-4}$ \\[0.1cm]
$x_{touch}+x_{susp}$ [m] & 618.7269 & 618.7099 & 0.0027 \\[0.1cm]
$z_{top}$ [m] & 71.2 & 71.0273 & 0.2426 \\[0.1cm]
\hline\hline
\end{tabular}
\end{table}

\begin{table}[H]
\centering
\caption{ROD-B (case 6) - Force control.}
\label{Case6.TA4}
\begin{tabular}{ >{\arraybackslash}m{.35\linewidth} 
>{\centering\arraybackslash}m{.14\linewidth} 
>{\centering\arraybackslash}m{.18\linewidth} >{\centering\arraybackslash}m{.14\linewidth}}
\hline\hline
Solution & Ref. sol. \cite{Gunnar2024}  & ARMoor ($N_e=128$)  & $\overline{\text{Diff.}}_X(\cdot,\cdot)$ $[\%]$\\ 
\hline
$x_{touch}$ [m] & 86.17 & 85.1198 & 1.218  \\ [0.1cm]
$\theta(L)$ [$^{\circ}$] & 14.9079 & 14.9000 & 0.053 \\[0.1cm]
Stretched length, $L_f$ [m] & 630.55 & 630.5619 & 0.0019 \\[0.1cm]
$x_{touch}+x_{susp}$ [m] & 624.3158 & 624.3112 & 7.36$\times 10^{-4}$ \\[0.1cm]
$z_{top}$ [m] & 71.2 & 70.9959 & 0.286 \\[0.1cm]
\hline\hline
\end{tabular}
\end{table}


\section*{Acknowledgments}
C.G. Gebhardt, B.A. Roccia and T.-H. Nguyen gratefully acknowledge the financial support from the European Research Council through
the ERC Consolidator Grant “DATA-DRIVEN OFFSHORE” (Project ID 101083157).

\bibliographystyle{unsrt}
\bibliography{Mybib}   

\end{document}